\documentclass[fleqn]{rbfin}
\usepackage{soul}
\usepackage{xcolor}
\usepackage{enumitem}
\usepackage{pgfplots}
\pgfplotsset{compat=1.18}
\usepackage[labelfont=bf,labelsep=colon]{caption}
\begin{document}
% SELECT ARTICLE LANGUAGE
% Para artigos em PORTUGUÊS, alterar a linha abaixo para \selectlanguage{brazilian}
\selectlanguage{english}
\raggedbottom
\setlength{\emergencystretch}{2em}
\hfuzz=3pt
% Set the highlight color to blue
\sethlcolor{cyan}
\frenchspacing
\setlist[itemize]{noitemsep, topsep=2pt}
\setlist[itemize]{noitemsep, topsep=2pt}

%%%%%%%%%%%%%%%%%%%%%%%%%%%%%%%%%%%%
% FRONT MATTER
%%%%%%%%%%%%%%%%%%%%%%%%%%%%%%%%%%%%

% ARTICLE TITLE
\title{Exploring Student Perception on Gen AI Adoption in Higher Education: A Descriptive Study} % Appears on title page

%%%%%%%%%%%%%%%%%%%%%%%%%%%%%%%%%%%%
% AUTHOR INFORMATION
%%%%%%%%%%%%%%%%%%%%%%%%%%%%%%%%%%%%
\author{\centering
% Authors in 2 rows x 3 columns
\begin{tabular}{@{}C{0.31\linewidth}C{0.31\linewidth}C{0.31\linewidth}@{}}
\textbf{Harpreet Singh}\orcidlink{0000-0001-6039-6834}
&
\textbf{Jaspreet Singh}%\orcidlink{}
&
\textbf{Satwant Singh}%\orcidlink{}
\\[-1pt]
\scriptsize\href{mailto:h.singh@tees.ac.uk}{h.singh@tees.ac.uk}
&
\scriptsize\href{mailto:J.singh@tees.ac.uk}{J.singh@tees.ac.uk}
&
\scriptsize\href{mailto:singh.satwant61@yahoo.in}{singh.satwant61@yahoo.in}
\\[7pt]
\textbf{Rupinder Singh}\orcidlink{0000-0001-6530-9453}
&
\textbf{Shamim Ibne Shahid}%\orcidlink{}
&
\textbf{Mohammad Hassan Tayarani Najaran}\orcidlink{0000-0002-5999-2134}
\\[-1pt]
\scriptsize\href{mailto:rupinder@outlook.in}{rupinder@outlook.in}
&
\scriptsize\href{mailto:s.shahid2@herts.ac.uk}{s.shahid2@herts.ac.uk}
&
\scriptsize\href{mailto:m.tayaraninajaran@herts.ac.uk}{\nolinkurl{m.tayaraninajaran@herts.ac.uk}}
\end{tabular}
}

\maketitle

\begin{abstract}
The rapid proliferation of Generative Artificial Intelligence (GenAI) is reshaping pedagogical practices and assessment models in higher education. While institutional and educator perspectives on GenAI integration are increasingly documented, the student perspective remains comparatively underexplored. This study examines how students perceive, use, and evaluate GenAI within their academic practices, focusing on usage patterns, perceived benefits, and expectations for institutional support. Data were collected through a questionnaire administered to 436 postgraduate Computer Science students at the University of Hertfordshire and analysed using descriptive methods. The findings reveal a \textit{Confidence--Competence Paradox}: although more than 60\% of students report high familiarity with tools such as ChatGPT, daily academic use remains limited and confidence in effective application is only moderate. Students primarily employ GenAI for cognitive scaffolding tasks, including concept clarification and brainstorming, rather than fully automated content generation. At the same time, respondents express concerns regarding data privacy, reliability of AI-generated information, and the potential erosion of critical thinking skills. The results also indicate strong student support for integrating AI literacy into curricula and programme Knowledge, Skills, and Behaviours (KSBs). Overall, the study suggests that universities should move beyond a \textit{policing} approach to GenAI and adopt a \textit{pedagogical} framework that emphasises AI literacy, ethical guidance, and equitable access to AI tools.

% KEYWORDS (separated by semicolon)
\keywordslist{AI and education (AIeD), GenAI in education, AI in Higher education}
% JEL CODES (separated by comma)
%\jelcodeslist{E3, C41, C43.}
\end{abstract}

%%%%%%%%%%%%%%%%%%%%%%%%%%%%%%%%%%%%
% END OF FRONT MATTER
%%%%%%%%%%%%%%%%%%%%%%%%%%%%%%%%%%%%

%%%%%%%%%%%%%%%%%%%%%%%%%%%%%%%%%%%%
% DOCUMENT MAIN SECTIONS
%%%%%%%%%%%%%%%%%%%%%%%%%%%%%%%%%%%%

\section{Introduction}\label{sec-intro}

The rise of generative artificial intelligence (GenAI) and various tools (e.g., ChatGPT, Genmini, Claude) over the past few years has marked a transformative moment in the history of higher education. The technological advancements have transformed higher education, moving from physical textbooks to digital resources, the rise of internet enabled online learning platforms, smartphones and apps enabled flexible learning. Today, GenAI systems such as ChatGPT extend this trajectory by introducing tools that not only mediate but also co-create knowledg. Thus, it becomes very important to study and understand the impact of GenAI within the broader context of digital transformation, examining how students as primary stakeholders, perceive and navigate its growing influence.

\subsection{Digital Transformation in Higher Education}
Higher education has gone through four major stages of digital transformation. The first stage began in the 1990s, when internet access and digitised resources expanded online teaching and learning. Learning Management Systems (LMSs), such as Blackboard and Moodle, became central tools for content delivery, communication, and course management ~\cite{AlEmran2018Investigating, Brynjolfsson2014SecondMachine}. The second stage followed with Web 2.0 and MOOCs, where platforms like Coursera and edX widened access and encouraged network-based, participatory learning ~\cite{Hollands2014MOOCs, Siemens2005Connectivism}. The third stage appeared in the 2010s through early AI applications, including adaptive learning and automated feedback, but it also raised concerns about bias, fairness, and educator autonomy ~\cite{Ifenthaler2020Utilising, Williamson2020Historical}. The current fourth stage started with ChatGPT in 2022 and continued with tools such as Gemini, Claude, and DeepSeek, which can generate original text and images and have intensified debates on authorship, originality, and academic integrity ~\cite{Hu2023ChatGPT, Luo2024AcademicIntegrity, Kasneci2023ChatGPT}.

\subsection{AI as a Disruptive Force}  
Generative AI (GenAI) is a subclass of AI that enables the development of models capable of generating original and contextually relevant content such as text, images, and multimedia ~\cite{Brown2020Language}. These models are trained on vast amounts of data from the Internet using large-scale computational infrastructure, including high-performance data centres. GenAI has become a disruptive force in education due to its capability to provide personalized tutoring, instant explanations, and adaptive feedback to learners. It also supports multilingual learners by providing accessible materials ~\cite{Ng2024Equity}, while educators benefit from assistance in rubric generation, assessment design, and content summarization ~\cite{Sousa2025GenerativeAI}. However, these capabilities also introduce risks. Academic integrity is challenged as conventional assessment models struggle to evaluate AI-generated work ~\cite{Ardito2024GenerativeAI}. Concerns about biased or inaccurate outputs also raise issues of reliability and misinformation ~\cite{Bender2021Dangers}. In addition, GenAI poses challenges for educators regarding job displacement and shifting professional roles ~\cite{Ivanov2024Drivers}. Institutional responses to GenAI remain uneven. Some universities have banned AI tools and classify their use as academic misconduct ~\cite{Heaven2023ChatGPT}, while others advocate their constructive and ethical integration in education ~\cite{RussellGroup2023Principles}. International organisations such as UNESCO and OECD have proposed human-centred governance frameworks that align innovation with accountability and transparency ~\cite{UNESCO2023Guidance, OECD2023DigitalEducation}.

\subsection{Expectations and Institutional Responsibilities} 
Following the disruptive dynamics discussed in Section 2.2, student expectations are becoming more specific. Many students now treat GenAI as part of normal academic and professional practice, and they expect clear, consistent guidance on acceptable use ~\cite{Kelly2023Generative, Arowosegbe2024Students}. When institutional rules are unclear, students report anxiety about misconduct and uncertainty about boundaries ~\cite{Kelly2023Generative, Arowosegbe2024Students, Yusuf2024GenerativeAI}.

Students generally favour guided integration rather than strict prohibition. Evidence indicates support for ethical boundaries, transparent policy communication, and practical direction on how GenAI can be used within disciplines ~\cite{Luo2024CriticalReview, Weng2024Assessment}. This expectation also extends to curriculum design: students increasingly want AI literacy and wider digital competence embedded as core employability skills ~\cite{Kelly2023Generative, Chiu2024FutureResearch}. Institutional responsibility therefore goes beyond policy statements. Students expect educators to act as facilitators, supported by workshops, seminars, and other structured forms of training that translate policy into practice ~\cite{Weng2024Assessment, Sousa2025GenerativeAI}. Overall, the literature suggests that effective governance depends on a balanced approach: clear rules, discipline-specific guidance, and practical learning support that reduces uncertainty while encouraging responsible use ~\cite{Luo2024CriticalReview, Yusuf2024GenerativeAI}.

Although research on generative artificial intelligence (GenAI) in education is expanding rapidly, several important gaps persist. Existing studies remain largely confined within national contexts, offering limited cross-regional or comparative perspectives that would enable broader generalization of findings ~\cite{Ivanov2024Drivers}. Moreover, much of the current scholarship privileges institutional and educator viewpoints, leaving student experiences and expectations underrepresented in analyses of GenAI adoption ~\cite{Weng2024Assessment, Luo2024CriticalReview}. Institutional frameworks addressing GenAI also tend to be fragmented, often focusing separately on ethical, pedagogical, or policy dimensions rather than integrating these into coherent, system-wide strategies ~\cite{UNESCO2023Guidance, RussellGroup2023Principles}. In addition, while innovative assessment models have been proposed to respond to GenAI’s impact on academic integrity, there remains a lack of empirical evidence validating their pedagogical effectiveness or student acceptability ~\cite{Roe2024GenAIFeedback}. Addressing these shortcomings calls for research that centers student perspectives, situates them within the historical evolution of educational technology, and develops actionable frameworks for institutional implementation. 
\subsection{Research Gap and Contribution }
% % This research contributes to the field of educational technology as well as artificial intelligence in education (AIeD) by providing empirical evidence of the \textit{Utility Gap} the discrepancy between students’ high perceived essentiality of GenAI and their relatively low frequency of actual academic use. Unlike previous studies that often treat students as a monolithic group of \textit{early adopters} this study identifies a \textit{false peak} of familiarity, where usage fluency masks gaps in critical competence and ethical application. Furthermore, by utilising both multiple-choice and open-text responses, this research identifies a \textit{social desirability bias} in student reporting: students underreport GenAI use for writing tasks when prompted directly but acknowledge its role close to a \textit{super-editor} in free-text formats. Finally, the study offers a timely analysis of the \textit{Algorithmic Paywall} highlighting how reliance on premium third-party tools creates new forms of digital inequality that universities must address through institutional provision.

Despite growing awareness of GenAI among university students, existing research provides limited empirical insight into how this awareness translates into actual academic practice. In this paper, we analyses survey data from 436 postgraduate Computer Science students at the University of Hertfordshire to examine how students perceive and use generative artificial intelligence (GenAI) in academic contexts. This study makes the following contributions: 
\begin{itemize}

\item It identifies a \textit{Utility Gap}, showing a mismatch between students’ perception of GenAI as essential and their relatively low frequency of academic use.

\item The findings reveal a \textit{False Peak of Familiarity}, where apparent fluency with GenAI masks gaps in critical competence and ethical understanding.

\item Using both multiple-choice and open-text responses, the study uncovers a \textit{Social Desirability Bias} in reporting AI-assisted writing practices.

\item It also highlights an \textit{Algorithmic Paywall}, where reliance on premium AI tools may create new forms of digital inequality in higher education.
\end{itemize}

\section{Literature review}
\subsection{Awareness and Exposure}
Prior research on tertiary-level students’ awareness of AI suggests that over 80\% are familiar with tools such as ChatGPT; however, deeper literacy in responsible and critical use remains limited ~\cite{MitevskaPetrusheva2023AITechnologies}. Similarly, although awareness appears high across disciplines, students often overestimate their competence, particularly in applying GenAI tools ethically ~\cite{Kelly2023Generative}. This mismatch between familiarity and responsible engagement remains a central theme in recent research.

Recent studies ~\cite{Xia2024ScopingReview} show that students frequently use GenAI for brainstorming, drafting, and feedback, yet often struggle to evaluate output accuracy. The same pattern appears in the U.S. ~\cite{Klimova2025UseOfChatGPT}, where ChatGPT is widely used for understanding complex concepts and structuring assignments, but concerns persist about overreliance and reduced critical thinking. Evidence from the UAE further suggests that adoption is shaped by trust, effort expectancy, and hedonic motivation ~\cite{Shuhaiber2025ChatGPT}. Across disciplines, use cases vary (e.g., coding in STEM, writing support in humanities), and perceptions of reliability also differ.

A multicultural view indicates that awareness is high globally, but exposure is context-dependent ~\cite{Yusuf2024GenerativeAI}. In Southeast Asia, students report lower confidence in independent use and prefer blended approaches where AI feedback is mediated by human guidance ~\cite{Roe2024GenAIFeedback}. In Hong Kong, students recognize personalization benefits but remain cautious about creativity and privacy ~\cite{Chan2024GenerativeAI}. A multi-university policy analysis similarly shows stronger guideline development in the Global North, while Global South institutions face equity and infrastructure constraints ~\cite{Jin2025GenerativeAI}.

Studies capturing student voices also show both enthusiasm and caution. UK students report broad awareness and use for academic writing but uncertainty about ethical boundaries and institutional expectations ~\cite{Arowosegbe2024Perception}. U.S. students report frequent use for clarification tasks alongside concerns about privacy and reduced independence ~\cite{Klimova2025UseOfChatGPT}. This rapid normalization of GenAI in everyday study practices ~\cite{Kurtz2024Strategies} can, without deliberate strategy, widen inequality and encourage instrumental rather than critical use ~\cite{Francis2025GenerativeAI}.

The ethical dimension is increasingly central. Heightened awareness without guidance risks normalizing AI-assisted plagiarism and weakening academic integrity ~\cite{Kovari2025EthicalUse}. Reported concerns include plagiarism, superficial learning, and diminished creativity ~\cite{Klimova2025UseOfChatGPT}, while perceived risk also shapes adoption intentions, reinforcing the need for transparency and institutional trust-building ~\cite{Shuhaiber2025ChatGPT}. Effective exposure therefore requires digital literacies such as fact-checking, bias recognition, and critical engagement ~\cite{Kasneci2023ChatGPT}.

The institutional environment plays a pivotal role. Framing GenAI primarily as a threat to originality can limit constructive engagement ~\cite{Luo2024CriticalReview}. By contrast, many European and U.S. universities are moving toward balanced guidance that permits use while stressing AI literacy and ethics ~\cite{Weng2024Assessment,ChristBrendemuhl2025Leveraging}. Faculty-focused evidence supports this direction: many teachers support integration but call for clearer guidance, and German guidelines increasingly frame GenAI opportunities as outweighing risks ~\cite{Bender2021Dangers,ChristBrendemuhl2025Leveraging}. Taken together, awareness is widespread, but responsible exposure depends on proactive literacy initiatives, scaffolded policy, and cross-disciplinary support.

\subsection{AI Literacy in the Curriculum}
As GenAI becomes increasingly embedded in higher education, students and educators alike recognise AI literacy as a fundamental component of future-ready curricula. AI literacy extends beyond technical skills and includes ethical, critical, and creative dimensions that support meaningful engagement with AI tools. It has been described as an “essential digital literacy” for navigating a world where intelligent technologies increasingly mediate learning and professional practice~\cite{Bender2024Awareness}. Ethical and responsible AI use also requires evaluative judgment, bias detection, and critical interpretation of AI-generated content~\cite{Yusuf2024GenerativeAI}. Without explicit instruction in these areas, students may over-rely on AI or misuse it in ways that undermine learning integrity.

Empirical studies therefore highlight the importance of embedding AI literacy into formal instruction rather than treating it as optional. In a large-scale Australian survey of 1,135 university students, most respondents reported limited confidence in using GenAI tools and called for structured learning activities to build these competencies~\cite{Kelly2023Generative}. The study also argues that students should be explicitly taught appropriate use of generative AI through discipline-specific learning activities. Evidence from Hong Kong similarly shows that students value AI for idea generation and personalised learning, while expecting educators to provide explicit guidance on responsible use~\cite{Chan2023StudentsVoices}.

Practical frameworks have also been proposed, including the AI Assessment Integration Framework and the Six Assessment Redesign Pivotal Strategies (SARPS), to embed AI literacy outcomes into teaching, learning, and assessment. Developing AI literacy in curricula also aligns with institutional and policy priorities. Both educators and students need support through professional development to ensure consistent understanding of AI tools, their ethical implications, and pedagogical integration~\cite{Ayyoub2025Advancing}. From a policy perspective, AI literacy is also positioned as a core learning outcome in higher education, with emphasis on equitable access to training and a culture of critical engagement~\cite{Chan2024GenerativeAI}. Overall, integrating AI literacy across curricula helps students become informed, ethical, and innovative thinkers who can co-work with AI in academic and professional contexts.

\subsection{Student Expectations of Educators and Institutions in GenAI Integration}
The rise of generative GenAI in higher education (HE) has reshaped what students expect from educators and institutions. Students consistently call for explicit, transparent, and fair policies on AI use, because ambiguity increases anxiety around academic misconduct ~\cite{Arowosegbe2024Students}. This concern is echoed in policy reviews showing that universities often frame GenAI as a threat to originality without fully addressing its potential as a collaborative academic tool ~\cite{Luo2024CriticalReview}.

Students therefore expect more nuanced guidance that distinguishes unethical use from legitimate academic support. Evidence suggests that students value discipline-specific rules and accessible support such as workshops and consultations ~\cite{Weng2024Assessment}. They also expect AI to be integrated into curricula and assessment rather than excluded: students in Hong Kong and Australia report valuing AI for brainstorming and personalised support, while still expecting assessment redesign that protects fairness, integrity, creativity, and critical thinking ~\cite{Chan2023StudentsVoices,Kelly2023Generative}. In this context, GenAI is seen as potentially supporting self-regulated learning, but only when institutions redesign assessment in line with these new affordances ~\cite{Xia2024ScopingReview,Weng2024Assessment}. Students likewise expect a shift from easily automated traditional tasks toward authentic and interdisciplinary assessment for AI-rich workplaces ~\cite{Chiu2024GenerativeAI,Sousa2025GenerativeAI,Khlaif2025Redesigning}.

Another central expectation is structured development of AI literacy. Students emphasize that universities should provide formal opportunities to learn responsible AI use, not just tool familiarity. AI literacy is framed as an essential digital literacy spanning technical, ethical, and critical dimensions, including evaluative judgment, bias detection, and critical interpretation of outputs ~\cite{Bender2024Awareness,Yusuf2024GenerativeAI}. Conceptual and policy work similarly stresses integrating AI literacy across higher education through curriculum and assessment frameworks, alongside equitable access to training ~\cite{Chan2024GenerativeAI}.

Students also expect institutions to reduce inequity in access. Reliance on commercial AI platforms can deepen digital divides, especially where cost and infrastructure constraints limit adoption. Comparative evidence shows that students expect universities to play an equalising role by providing institutional licences, subsidised access, and support tailored to varied levels of digital competence ~\cite{Sousa2025GenerativeAI,Zubair2025Determinants}. Without such measures, AI integration risks widening existing disparities.

At the same time, students do not view AI as a replacement for educators. They expect educators to act as mentors who model responsible and critical AI use. Evidence from Vietnam and Singapore suggests that students value AI-generated feedback most when paired with instructor commentary, indicating a preference for a human--AI partnership ~\cite{Roe2024GenAIFeedback}. Related work also notes expectations that educators demonstrate ethical engagement with AI, supported by professional development for staff in AI-rich contexts ~\cite{Francis2025GenerativeAI,Ayyoub2025Advancing}.

Students further expect institutions to protect academic integrity through constructive and transparent systems rather than purely punitive enforcement. Many students recognise plagiarism and misuse risks, but remain sceptical about detector fairness and reliability ~\cite{Ardito2024GenerativeAI,Roe2024GenAIFeedback}. Concerns include inconsistency, possible bias against non-native English writers, and vulnerability to circumvention tools ~\cite{Ogunleye2024HigherEducation,singh2023exploring}.

Overall, students expect institutions to provide clear policy, fair assessment reform, educator readiness, ethical guidance, and equitable access. Their expectations point to supportive governance that enables responsible use while preserving academic integrity ~\cite{Weng2024Assessment,Chiu2024GenerativeAI}.

\subsection{Institutional role in the integration of AI tools in Higher Education.}
The emergence of GenAI has required universities to move quickly from ad hoc responses to coordinated institutional strategies that combine governance, staff support, and pedagogical redesign ~\cite{Nikolic2024Systematic, Weng2024Assessment, Xia2024ScopingReview}. Because GenAI can complete many conventional written tasks, higher education institutions face growing pressure to update academic-integrity frameworks without reducing policy to prohibition alone ~\cite{Luo2024CriticalReview, Rudolph2023ChatGPT, Weng2024Assessment}.

At policy level, the literature shows that many institutional rules remain too generic for current GenAI use cases ~\cite{Luo2024CriticalReview, Nikolic2024Systematic}. Universities often frame GenAI primarily as a threat to originality, but this framing can overlook collaborative and technology-mediated forms of contemporary knowledge production ~\cite{Luo2024CriticalReview, Wang2024GenerativeAI}. More effective approaches combine clear disclosure and attribution requirements with culturally sensitive implementation, supported by integrated governance models such as the AI Ecological Education Policy Framework ~\cite{Chan2023StudentsVoices, Khlaif2024UniversityTeachers, Weng2024Assessment, Yusuf2024GenerativeAI}.

Beyond governance, institutions must invest in educator development and assessment reform. Evidence indicates that teaching staff often lack sufficient training, particularly in AI literacy, assessment literacy, and the ethical limitations of GenAI outputs ~\cite{Ayyoub2025Advancing, Kurtz2024Strategies, Nikolic2024Systematic, Wang2024GenerativeAI}. In parallel, curricula and assessment need to shift toward higher-order outcomes and authentic tasks, including approaches where students critique AI-generated outputs, while avoiding overreliance on unreliable AI-detection systems ~\cite{Francis2025GenerativeAI, Khlaif2024UniversityTeachers, Khlaif2025Redesigning, Perkins2024GenAIDetection, Weng2024Assessment, Xia2024ScopingReview}.

%\subsection{The Research Gap}
%\textcolor{blue}{Focus on Student Perception
%Although student awareness of GenAI is now widespread, existing research offers limited empirical detail on how students translate this awareness into day-to-day academic practice, and what forms of institutional support they expect in response. In particular, the literature rarely quantifies the gap between perceived familiarity and actual academic use, nor does it clearly explain why confidence and effective application remain only moderate despite high self-reported exposure. Evidence also remains fragmented on the student-side drivers of adoption including perceived risks, ambiguity around acceptable practice, and access constraints and on how these factors relate to students' demands for AI literacy training, assessment guidance, and equitable access to tools. 
%To address this gap, this study provides a descriptive analysis of 436 postgraduate Computer Science students at the University of Hertfordshire, examining (i) patterns of GenAI familiarity and academic use, (ii) confidence and perceived competence, (iii) perceived risks and negative impacts, and (iv) expectations for lecturer guidance, institutional policy communication, curriculum integration, and equal access.} 

\section{Methodology}
The data for this study was collected from students through a questionnaire on institutional AI policy. The participants were students enrolled in the May 2025 session at the University of Hertfordshire under the module \textit{Research Methods in Computer Science}. The data were collected from 436 Computer Science students.

\noindent\textbf{Overview of moduel Research Methods in Computer Science:}
This is a postgraduate-level module, weighted at 30 credits, focusing on Research Methods in Computer Science. The teaching structure comprises a two-hour lecture session delivered weekly over an academic term of ten weeks.

\noindent\textbf{Learning Outcomes:}
Upon successful completion of this module, students will be able to:
\begin{itemize}
\item Discern and classify a comprehensive spectrum of research methodologies applicable to complex issues within the domain of computer science.
\item Comprehend and contextualise the practical application of these methods, specifically in relation to an advanced Master’s dissertation or project.
\item Critically appraise and implement a diverse set of planning and execution strategies that are essential for undertaking a substantial, self-directed research program at the advanced master’s level.
\end{itemize}

\noindent\textbf{Evaluation and Intervention:}
The intervention utilised was a multiple-choice, unmarked quiz administered via the Canvas learning management system. The core component of this intervention was a questionnaire designed to capture student familiarity, perspectives and expectations on institutional AI policy.

\noindent\textbf{Questionnaire Design:}
The instrument comprised 27 questions targeting various dimensions of student interaction with the GenAI tool. These aspects included:
\begin{itemize}
\item Students' familiarity with GenAI tools and patterns of academic use.
\item Self-reported confidence in using GenAI effectively, as a proxy for perceived competence.
\item Perceived benefits and risks of GenAI use in academic work.
\item Expectations for assessment-related guidance.
\item Perceptions of institutional communication and policy clarity regarding GenAI.
\item Preferences for curriculum integration and training provision.
\item Perceived consistency of staff engagement and encouragement to explore GenAI across modules.
\item Views on equity and access to tools.
\end{itemize}

\noindent\textbf{Response Format:} 
Most items used closed-response formats (e.g., multiple-choice selections and 5-point Likert-type agreement/satisfaction scales), supplemented by selected open-text prompts (labelled as '.b' sub-questions) to capture qualitative explanations.

\noindent\textbf{Demographic Variable:} The study intentionally excluded demographic variables to obtain a perception of GenAI and institutional policies that is agnostic to personal characteristics. This approach was facilitated by the controlled student group, as all participants are enrolled in the same M.Sc. program.\\
\noindent\textbf{Full Instrument:} The complete questionnaire is provided in Appendix A.\\
\noindent\textbf{Data Collection}
Data collection was conducted through the Canvas VLE. While the survey contained 27 questions, the platform's format necessitated splitting multi-part questions into individual numbered entries. To maintain clarity in Section IV, follow-up qualitative prompts are identified by the suffix '.b' and paired with their primary question. As a result, certain numerical gaps exist in the sequence ( 3, 6, 17, 21 and 27) to account for these integrated question pairs.\\
\textbf{Anonymity and Confidentiality:} To ensure candid and unhesitating responses, the questionnaire was fully anonymised. No demographic data that could potentially reveal the individual identity of any student was collected.\\
\textbf{Data Processing:} Following collection, the raw data were exported to a comma-separated values (CSV) file. Python was then utilised to conduct the required statistical analysis.\\
\textbf{Sample Size:} The analysis was focused on the responses received from all 436 students who participated in the study.
\section{Survey Result and Analysis}
This section presents the main findings from the student survey on generative AI use in higher education. We first report the descriptive results for each question and then interpret what these patterns suggest about student familiarity, confidence, concerns, and expectations of institutional support. Together, these results highlight key areas where universities can strengthen AI literacy, guidance, and equitable access.

\noindent\textbf{Q1: How familiar are you with generative AI tools?}

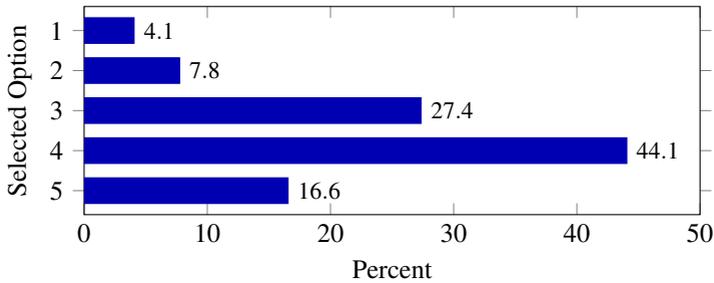
\begin{figure}[t]
\centering
\begin{tikzpicture}
\begin{axis}[
    xbar,
    width=0.9\linewidth,
    height=0.4\linewidth,
    xmin=0,
    xmax=50,
    bar width=10pt,
    enlarge y limits=0.15,
    xlabel={Percent},
    ylabel={Selected Option},
    symbolic y coords={1,2,3,4,5},
    ytick=data,
    y dir=reverse,
    nodes near coords,
    every node near coord/.append style={anchor=west,font=\small}
]

\addplot[draw=none,fill=blue!70!black]
coordinates {
    (4.1,1)
    (7.8,2)
    (27.4,3)
    (44.1,4)
    (16.6,5)
};

\end{axis}
\end{tikzpicture}
\caption{Student familiarity with generative AI tools on a 1–5 scale, where 1 represent no familiarity and 5 represent extreme familiarity.}
\label{fig:Q1}
\end{figure}

Figure \ref{fig:Q1} shows that over 60\% of respondents reported being “Very” or “Extremely” familiar with generative AI. At first glance, this suggests a digitally native group that has easily adopted these tools. However, this high level of reported familiarity contrasts with the moderate confidence levels expressed in later questions. This gap indicates that many students may be confusing “usage familiarity,” such as the ability to prompt a chatbot, with “critical familiarity,” which involves understanding limitations like hallucinations and bias. The data therefore points to a “False Peak” ~\cite{ kim2025examining} of familiarity among students, where self-reported fluency likely hides gaps in practical and critical competence.
In addition, the 27.4\% of students who identified themselves as “Moderately familiar” form an important risk group. These students possess sufficient knowledge to utilise generative AI for academic tasks, but may lack the deeper understanding required to use it responsibly. From a university perspective, this overconfidence is concerning ~\cite{mah2024artificial} because it can lead to complacency. Students who believe they already understand AI well are less likely to participate in voluntary training, which can create a competence trap where institutions assume a level of skill that is not actually present. Therefore, it is important for universities to help students accurately recognize their own level of familiarity with AI, as this awareness can encourage them to engage in training and capacity-building initiatives.\\

\newcommand{\QTwoLeftBlockWidth}{0.58\linewidth}
\newcommand{\QTwoRightBlockWidth}{0.41\linewidth}

\newcommand{\QTwoLeftImgHeight}{0.24\textheight}
\newcommand{\QTwoRightImgHeight}{0.19\textheight}

\noindent
\begin{minipage}[t]{\QTwoLeftBlockWidth}
  \centering
  \parbox[c][\QTwoLeftImgHeight][c]{\linewidth}{%
    \centering
    \resizebox{\linewidth}{\QTwoLeftImgHeight}{%
      \begin{tikzpicture}
\begin{axis}[
    xbar,
    width=\linewidth,
    height=0.27\textheight,
    xmin=0,
    xmax=40,
    bar width=10pt,
    enlarge y limits=0.12,
    xlabel={Percent},
    ylabel={Selected option},
    symbolic y coords={Other,None,Deepseek,Copilot,Gemini,Claude,ChatGPT},
    ytick=data,
    nodes near coords,
    every node near coord/.append style={
        anchor=west,
        font=\small
    },
    point meta=explicit symbolic,
    xmajorgrids=false,
    ymajorgrids=false,
    axis line style={black},
    tick style={black},
]
\addplot[
    draw=none,
    fill=blue!70!black
] coordinates {
    (2.7,Other)    [2.7]
    (0.1,None)     [0.1]
    (12.7,Deepseek)[12.7]
    (20.4,Copilot) [20.4]
    (23.5,Gemini)  [23.5]
    (7.3,Claude)   [7.3]
    (33.3,ChatGPT) [33.3]
};
\end{axis}
\end{tikzpicture}%
    }%
  }

  \vspace{4pt}
  \captionsetup{hypcap=false}
  
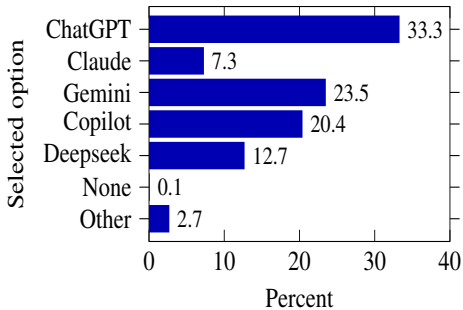
\captionof{figure}{Various GenAI tools that students are familiar with.}
  \label{Q2}
\end{minipage}\hfill
\begin{minipage}[t]{\QTwoRightBlockWidth}
  \centering
  \parbox[c][\QTwoLeftImgHeight][c]{\linewidth}{%
    \centering
    \raisebox{0.05\textheight}{%
      \fbox{%
        \includegraphics[width=0.9\linewidth,height=\QTwoRightImgHeight]{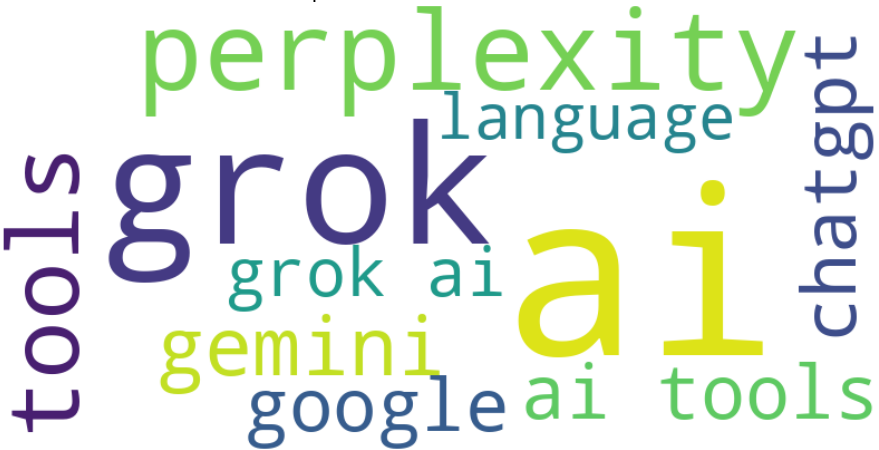}%
      }%
    }%
  }

  \vspace{6pt}
  \captionsetup{hypcap=false}
  \captionof{figure}{Word cloud of generative AI tools identified by students in text response.}
  \label{Q2.b}
\end{minipage}

\noindent\textbf{Q2 \& Q2(b): Which Gen AI tools do you have experience with?}

Figure \ref{Q2} shows that ChatGPT is the most widely used generative AI tool, reported by 33.3\% of respondents. This is followed by Gemini at 23.5\% and Copilot at 20.4\%. Smaller but still notable shares of students report using DeepSeek at 12.7\% and Claude at 7.3\%, while only a very small number indicate no experience with generative AI tools. This pattern suggests that students are not relying on a single dominant system but are instead using multiple AI platforms. Their choices often appear to be shaped by specific academic tasks or by how well the tools are integrated into familiar software environments. The accompanying word cloud in Fig. \ref{Q2.b} supports this interpretation, showing frequent references to AI tools that are embedded within larger ecosystems such as Google and Microsoft, as well as the growing visibility of answer-engine platforms like Perplexity and Grok. The strong presence of these externally adopted tools points to the rise of Shadow IT, where students independently use AI systems outside institutional provision or oversight. These findings indicate a fragmented AI usage landscape in which students act as multi-model users. From an institutional perspective, this fragmentation makes standardisation more difficult and raises concerns about equity, as differences in model quality, access to paid features, and platform integration may lead to unequal academic advantages among students.\\
\textbf{Q4: How frequently do you use AI tools for academic purposes?}

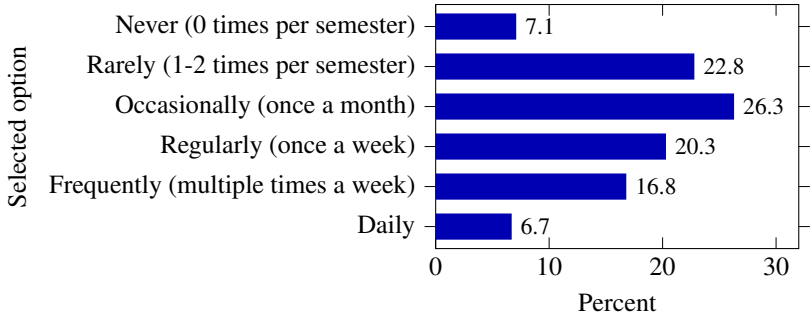
\begin{figure}[t]
\centering
\resizebox{\linewidth}{!}{\begin{tikzpicture}
\begin{axis}[
    xbar,
    width=0.6\linewidth,
    height=0.45\linewidth,
    xmin=0,
    xmax=32,
    bar width=10pt,
    enlarge y limits=0.11,
    xlabel={Percent},
    ylabel={Selected option},
    symbolic y coords={Daily,
    {Frequently (multiple times a week)},
    {Regularly (once a week)},
    {Occasionally (once a month)},
    {Rarely (1-2 times per semester)},
    {Never (0 times per semester)}},
    ytick=data,
    nodes near coords,
    every node near coord/.append style={
        anchor=west,
        font=\small
    },
    point meta=explicit symbolic,
    xmajorgrids=false,
    ymajorgrids=false,
    axis line style={black},
    tick style={black},
]
\addplot[
    draw=none,
    fill=blue!70!black
] coordinates {
    (6.7,Daily)  [6.7]
    (16.8,{Frequently (multiple times a week)}) [16.8]
    (20.3,{Regularly (once a week)}) [20.3]
    (26.3,{Occasionally (once a month)}) [26.3]
    (22.8,{Rarely (1-2 times per semester)}) [22.8]
    (7.1,{Never (0 times per semester)})  [7.1]
};
\end{axis}
\end{tikzpicture}}
\caption{Student frequency of GenAI use on for academic purposes.}
\label{Q4}
\end{figure}
 
Figure \ref{Q4} shows that only 6.7\% of respondents report daily academic use of AI tools. A further 16.8\% indicate frequent use, defined as multiple times per week, and 20.3\% report regular use (once a week). In contrast, a large share of students fall into lower-frequency categories, including occasional use at 26.3\%, rare use at 22.8\%, and no use at 7.1\%. Although awareness of AI tools is widespread, daily use remains limited, with most students using them only occasionally or rarely. This pattern challenges the common narrative of AI “addiction” and instead suggests that, for many students, AI functions as a problem-solving tool used only when needed rather than as a regular part of their academic routine. Such infrequent use is likely influenced by concerns about academic integrity ~\cite{pudasaini2024survey}. Students may deliberately limit their use to avoid detection by monitoring systems, which in turn prevents them from developing the practical skills needed for confident and responsible use of AI. From a pedagogical perspective, this reflects a lack of effective integration. If AI were embedded in teaching and learning as a tutor or research assistant, usage would be more consistent. Instead, the low frequency of use suggests that students view AI as a shortcut for specific challenges rather than as an ongoing support for learning. This points to a clear “Utility Gap” between the high level of familiarity students report and the low frequency ~\cite{kim2025examining, Arowosegbe2024Perception} with which they actually use these tools.\\

\newcommand{\QFiveLeftBlockWidth}{0.66\linewidth}
\newcommand{\QFiveRightBlockWidth}{0.34\linewidth}

\newcommand{\QFiveLeftImgHeight}{0.32\textheight}
\newcommand{\QFiveRightImgHeight}{0.2\textheight}

\noindent
\begin{minipage}[t]{\QFiveLeftBlockWidth}
  \centering
  \parbox[c][\QFiveLeftImgHeight][c]{\linewidth}{%
    \centering
    \resizebox{\linewidth}{\QFiveLeftImgHeight}{%
      \begin{tikzpicture}
\begin{axis}[
    xbar,
    width=0.5\linewidth,
    height=0.6\linewidth,
    xmin=0,
    xmax=85,
    bar width=5pt,
    enlarge y limits=0.08,
    xlabel={Percent},
    ylabel={Selected option},
    font=\tiny,
    ytick pos=left,
    symbolic y coords={
Other,
{Time-saving and efficiency improvement},
{Understanding or clarifying concepts},
{Mathematical computations and reasoning},
{Translation tasks},
{Summarizing texts or articles},
{Programming or coding tasks},
{Writing tasks (e.g., essays, reports)},
{Generating ideas or brainstorming}
},
    ytick=data,
    nodes near coords,
    every node near coord/.append style={
        anchor=west,
        font=\tiny
    },
]
\addplot[
    draw=none,
    fill=blue!70!black
] coordinates {
(2.8,Other) [2.8\%]
(47.2,{Time-saving and efficiency improvement}) [47.2\%]
(69.0,{Understanding or clarifying concepts}) [69.0\%]
(36.7,{Mathematical computations and reasoning}) [36.7\%]
(40.1,{Translation tasks}) [40.1\%]
(52.1,{Summarizing texts or articles}) [52.1\%]
(42.0,{Programming or coding tasks}) [42.0\%]
(36.9,{Writing tasks (e.g., essays, reports)}) [36.9\%]
(64.0,{Generating ideas or brainstorming}) [64.0\%]
};
\end{axis}
\end{tikzpicture}%
    }%
  }

  \vspace{4pt}
  \captionsetup{hypcap=false}
  
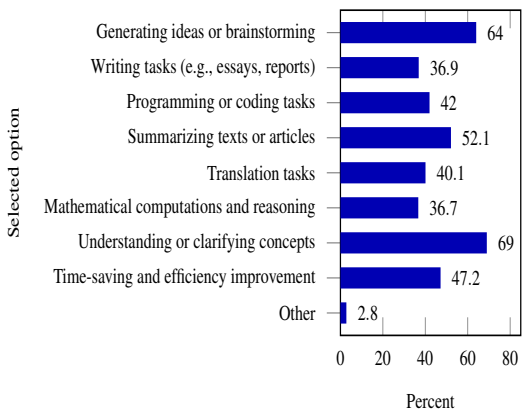
\captionof{figure}{Various ways AI support student learning and academic tasks.}
  \label{Q5}
\end{minipage}\hfill
\begin{minipage}[t]{\QFiveRightBlockWidth}
  \centering
  \parbox[c][\QFiveLeftImgHeight][c]{\linewidth}{%
    \centering
    \raisebox{0.05\textheight}{%
      \fbox{%
        \includegraphics[width=0.9\linewidth,height=\QFiveRightImgHeight]{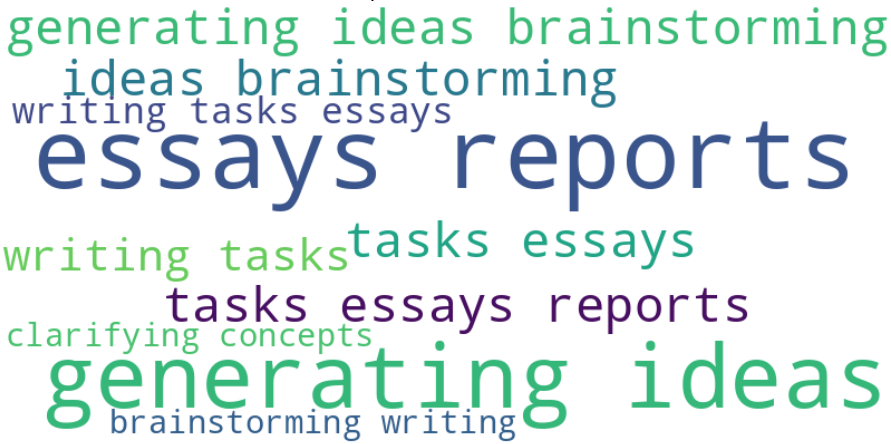}%
      }%
    }%
  }

  \vspace{6pt}
  \captionsetup{hypcap=false}
  \captionof{figure}{Word cloud of ways GenAI support students for academic tasks.}
  \label{Q5.b}
\end{minipage}

\noindent\textbf{Q5 \& Q5.(b): In what ways do AI tools support your learning?}

Reporting multiple responses, the Fig. \ref{Q5} shows that the most common uses of AI tools are related to understanding or clarifying concepts at 69.0\% and generating ideas or brainstorming at 64.0\%. These are followed by summarising texts or articles at 52.1\% and improving time efficiency at 47.2\%. Writing-related activities, such as essays or reports, are reported by a smaller share of respondents at 36.9\%. Similarly, programming or coding tasks are reported by 42.0\% of students, and translation tasks by 40.1\%. Overall, the data supports the “Cognitive Scaffolding” hypothesis and challenges the common assumption among faculty that students mainly use AI to generate essays from scratch. The high frequency of concept clarification and brainstorming suggests that students often use AI as a personal tutor to support learning and address gaps in formal instruction.
However, the open-text responses in Fig. \ref{Q5.b} present a more complex picture. In these responses, terms such as essays, reports, and writing tasks appear prominently. The difference between the multiple-choice results, where writing ranks lower, and the open-text responses suggests the presence of “Social Desirability Bias” in the survey. Students may underreport their use of AI for writing when asked directly, but are more open about it when given space to respond freely. This indicates that while cognitive scaffolding is the main reported use, a significant level of text generation is also taking place. Universities therefore need to clearly distinguish between acceptable practices such as editing and support, and unacceptable practices such as ghostwriting. Many students appear to view AI as a “Super-Editor” that helps refine their ideas, a distinction that is often overlooked in strict, zero-tolerance plagiarism policies.\\

\begin{minipage}[t]{0.49\linewidth}
\centering
\begin{tikzpicture}
\begin{axis}[
    xbar,
    width=1.0\linewidth,
    height=0.8\linewidth,
    xmin=0,
    xmax=50,
    bar width=10pt,
    enlarge y limits=0.12,
    xlabel={Percent},
    ylabel={Response},
    font=\scriptsize,
    ytick pos=left,
    symbolic y coords={5,4,3,2,1},
    ytick=data,
    nodes near coords,
    every node near coord/.append style={
        anchor=west
    },
]
\addplot[
    draw=none,
    fill=blue!70!black
] coordinates {
    (8.3,5)  [8.3]
    (26.7,4) [26.7]
    (36.2,3) [36.2]
    (19.1,2) [19.1]
    (9.7,1)  [9.7]
};
\end{axis}
\end{tikzpicture}

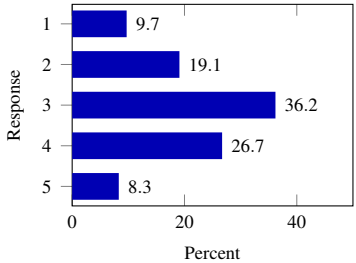
\captionof{figure}{Student confidence in using AI tools on a scale of 1–5, where 1 represents not at all confident and 5 represents extreme confidence.}
\label{Q7}
\end{minipage}\hfill
\begin{minipage}[t]{0.49\linewidth}
\centering
\begin{tikzpicture}
\begin{axis}[
    xbar,
    width=1.0\linewidth,
    height=0.8\linewidth,
    xmin=0,
    xmax=40,
    bar width=10pt,
    enlarge y limits=0.12,
    xlabel={Percent},
    ylabel={Response},
    font=\scriptsize,
    ytick pos=left,
    symbolic y coords={5,4,3,2,1},
    ytick=data,
    nodes near coords,
    every node near coord/.append style={
        anchor=west
    },
]
\addplot[
    draw=none,
    fill=blue!70!black
] coordinates {
    (11.7,5) [11.7\%]
    (28.0,4) [28.0\%]
    (30.1,3) [30.1\%]
    (22.9,2) [22.9\%]
    (7.3,1)  [7.3\%]
};
\end{axis}
\end{tikzpicture}

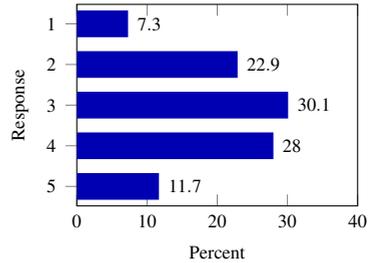
\captionof{figure}{Student view on GenAI being essential for their academic success on a scale of 1-5, where 1 represent not at all essential and 5 represents absolutely essential.}
\label{Q8}
\end{minipage}

\textbf{Q7. To what extent do you feel confident in your ability to use AI tools effectively?}\\
Figure \ref{Q7} shows that student confidence in using AI tools for academic tasks follows a bell-shaped pattern. The largest group of respondents report being Moderately confident at 36.2\%, followed by Very confident at 26.7\%. Smaller proportions identify as Slightly confident at 19.1\%, Not at all confident at 9.7\%, and Extremely confident at 8.3\%. When these confidence levels are compared with students’ views on the importance of AI for their future, a clear “Confidence-Competence Paradox” becomes visible. Although students see AI as essential, their confidence in using it effectively peaks at a moderate level ~\cite{kim2025examining}. This pattern reflects a state of “Conscious Incompetence” ~\cite{kim2025examining, asio2024ai}, where students are aware that their knowledge and skills are incomplete. This awareness can contribute to anxiety, as students feel pressure to master a technology that universities often describe as central to future academic and professional success, while receiving limited guidance on how to do so. From an institutional perspective, this finding signals an important opportunity. The dominance of moderate confidence suggests that students are realistic about their limitations and open to learning. However, without structured support, this uncertainty may lead to avoidance of AI tools or to hidden and potentially unethical forms of use.\\
\textbf{Q8: How essential are AI tools to your academic success?}

Figure \ref{Q8} shows that a large majority of respondents view generative AI tools as essential to their academic success. About 30.1\% describe them as Moderately essential, 28.0\% as Very essential, and 11.7\% as Absolutely essential. Together, these groups account for nearly 70\% of the sample. In contrast, smaller shares of students consider AI tools to be Slightly essential at 22.9\% or Not at all essential at 7.3\%. When majority of students categorize AI tools as essential, these systems move beyond the role of optional support and become part of core academic infrastructure ~\cite{luo2025design, liang2025systematic}, similar to internet access or digital learning platforms. Such reliance also creates vulnerability, as students become dependent on third-party corporate tools that may change in availability, cost, or performance over time. In addition, the perception of AI as “absolutely essential” suggests that many current assessment tasks focus on skills that AI can easily perform, such as basic summarisation or synthesis. This raises important concerns for curriculum design and assessment practices. If assessments primarily measure skills that are easily replicated by AI, students may feel compelled to rely on these tools to remain competitive. Universities therefore need to rethink assessment strategies to place greater emphasis on human-centered skills, including critical thinking, judgment, and originality, rather than on tasks that can be readily automated by AI systems.\\
\begin{minipage}[t]{0.49\linewidth}
\centering
\begin{tikzpicture}
\begin{axis}[
    xbar,
    width=1.0\linewidth,
    height=0.8\linewidth,
    xmin=0,
    xmax=50,
    bar width=10pt,
    enlarge y limits=0.12,
    xlabel={Percent},
    ylabel={Response},
    font=\footnotesize,
    ytick pos=left,
    symbolic y coords={5,4,3,2,1},
    ytick=data,
    nodes near coords,
    every node near coord/.append style={
        anchor=west
    },
]
\addplot[
    draw=none,
    fill=blue!70!black
] coordinates {
    (20.2,5) [20.2\%]
    (36.9,4) [36.9\%]
    (28.2,3) [28.2\%]
    (8.0,2)  [8.0\%]
    (6.7,1)  [6.7\%]
};
\end{axis}
\end{tikzpicture}

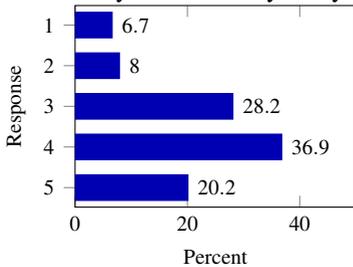
\captionof{figure}{Student view of resources and guidance University provided on GenAI on a scale of 1-5, where 1 represent strongly disagree and 5 represent strongly agree.}
\label{Q9}
\end{minipage}\hfill
\begin{minipage}[t]{0.49\linewidth}
\centering
\begin{tikzpicture}
\begin{axis}[
    xbar,
    width=1.0\linewidth,
    height=0.8\linewidth,
    xmin=0,
    xmax=55,
    bar width=10pt,
    enlarge y limits=0.12,
    xlabel={Percent},
    ylabel={Response},
    font=\footnotesize,
    ytick pos=left,
    symbolic y coords={5,4,3,2,1},
    ytick=data,
    nodes near coords,
    every node near coord/.append style={
        anchor=west
    },
]
\addplot[
    draw=none,
    fill=blue!70!black
] coordinates {
    (29.7,5) [29.7\%]
    (39.6,4) [39.6\%]
    (18.7,3) [18.7\%]
    (7.1,2)  [7.1\%]
    (4.8,1)  [4.8\%]
};
\end{axis}
\end{tikzpicture}

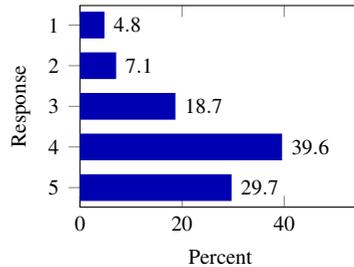
\captionof{figure}{Student-rated levels of awareness regarding University's official AI assessment policies on a scale of 1-5, where 1 represent not informed and 5 represent extremely informed.}
\label{Q10}
\end{minipage}

\noindent\textbf{Q9: My university has provided sufficient resources or guidance?}

Figure \ref{Q9} shows a mixed assessment of institutional support for effective use of AI. A total of 57.1\% of respondents agree that their university has provided sufficient resources or guidance, including 36.9\% who agree and 20.2\% who strongly agree. At the same time, a notable proportion of students express uncertainty or dissatisfaction. About 28.2\% select a neutral response, while 14.7\% report disagreement, including 8.0\% who disagree and 6.7\% who strongly disagree. This pattern suggests that although students are aware of institutional policies, many are not fully satisfied with the support provided to help them comply with those policies. The findings point to a “Policing over Pedagogy” approach ~\cite{pudasaini2024survey, mah2024artificial}, in which the institution has been effective in communicating restrictions on AI use but less effective in offering practical guidance on appropriate and productive use. The sizeable share of neutral and negative responses indicates that students clearly distinguish between being given permission and being given support. While policies and warnings are visible, students perceive a lack of investment in practical resources such as licensed tools, prompt libraries, and clear technical guidance. This gap between regulation and support risks creating a climate of frustration, where students feel that institutional efforts are focused more on managing risk and liability than on actively developing their AI-related skills for the future.\\
\textbf{Q10. How well-informed are you about official policies?}

Figure \ref{Q10} shows a high level of student awareness of institutional policies related to the use of AI in assessment. About 39.6\% of respondents report being Very informed, and a further 29.7\% identify as Extremely informed. Together, these groups account for 69.3\% of the sample. Smaller proportions describe themselves as Moderately informed at 18.7\%, Slightly informed at 7.1\%, or Not informed at all at 4.8\%. This unusually high level of policy awareness suggests that the university has been effective in communicating its rules on AI use. However, this apparent “Compliance Victory” ~\cite{mah2024artificial} may come at a cost. High awareness of restrictive policies often reflects a focus on control and surveillance rather than on learning and guidance. In this context, students may understand what is prohibited but receive little clarity on how AI can be used ethically and productively. This highlights a key distinction between notification and education. Knowing the consequences of misuse does not equip students with the skills needed to use AI responsibly. While the institution appears to have successfully managed risk, the findings suggest that it has not yet given equal attention to supporting innovation and ethical engagement with AI tools.\\

\begin{minipage}[t]{0.98\linewidth}
\centering
\begin{tikzpicture}
\begin{axis}[
    xbar,
    width=0.6\linewidth,
    height=0.4\linewidth,
    xmin=0,
    xmax=60,
    bar width=10pt,
    enlarge y limits=0.12,
    xlabel={Percent},
    ylabel={Response},
    font=\footnotesize,
    ytick pos=left,
    symbolic y coords={
        {Not aware of any such opportunities},
        {No, and I am not interested},
        {No, but I would be interested},
        {Yes, once},
        {Yes, multiple times}
    },
    ytick=data,
    nodes near coords,
    every node near coord/.append style={anchor=west},
]
\addplot[
    draw=none,
    fill=blue!70!black
] coordinates {
    (12.8,{Not aware of any such opportunities}) [12.8\%]
    (6.4,{No, and I am not interested}) [6.4\%]
    (48.9,{No, but I would be interested}) [48.9\%]
    (21.1,{Yes, once}) [21.1\%]
    (10.8,{Yes, multiple times}) [10.8\%]
};
\end{axis}
\end{tikzpicture}

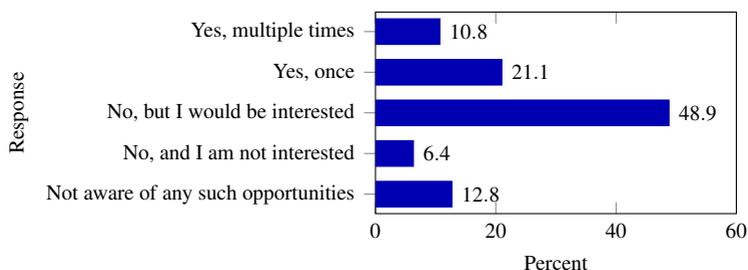
\captionof{figure}{Student participation levels in university-provided GenAI use in learning workshops.}
\label{Q11}
\end{minipage}\hfill

\textbf{Q11: Have you participated in any university-provided training?}

Figure \ref{Q11} shows that 32.1\% of respondents have taken part in university-provided training on generative AI, including 21.1\% who attended once and 10.8\% who participated multiple times. In contrast, 48.9\% of students report that they have not attended any training but would like to do so, while 12.8\% are not aware that such training exists. Only 6.4\% state that they are not interested in participating. These results reveal a clear “Latent Demand Crisis” ~\cite{kim2025examining, mah2024artificial, luo2025design}, marked by a wide gap between students who have received training and those who want access to it. This gap is unlikely to reflect low motivation, as interest levels are high across the sample. Instead, it suggests problems related to accessibility, visibility, or scheduling of training opportunities. When institutional support does not meet student demand, learners may turn to informal online sources for guidance, where information is often unregulated and may encourage unethical practices. The findings, therefore, indicate a missed opportunity for universities to shape responsible and effective AI use. With a large group of students actively seeking guidance, improving the reach and design of AI training programs represents a practical and immediate step toward strengthening academic support and skill development.\\

\begin{minipage}[t]{0.49\linewidth}
\centering
\begin{tikzpicture}
\begin{axis}[
    xbar,
    width=\linewidth,
    height=0.7\linewidth,
    xmin=0,
    xmax=55,
    bar width=10pt,
    enlarge y limits=0.12,
    xlabel={Percent},
    ylabel={Response},
    font=\footnotesize,
    ytick pos=left,
    symbolic y coords={5,4,3,2,1},
    ytick=data,
    nodes near coords,
    every node near coord/.append style={anchor=west},
]
\addplot[
    draw=none,
    fill=blue!70!black
] coordinates {
    (16.8,5) [16.8\%]
    (47.0,4) [47.0\%]
    (24.4,3) [24.4\%]
    (6.0,2)  [6.0\%]
    (5.8,1)  [5.8\%]
};
\end{axis}
\end{tikzpicture}

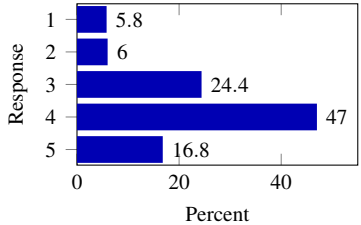
\captionof{figure}{Student agreement regarding the explicit inclusion of AI skills in program KSBs on scale of 1-5, where 1 represent strongly disagree and 5 represent strongly agree.}
\label{Q12}
\end{minipage}\hfill
\begin{minipage}[t]{0.49\linewidth}
\centering
\begin{tikzpicture}
\begin{axis}[
    xbar,
    width=\linewidth,
    height=0.7\linewidth,
    xmin=0,
    xmax=55,
    bar width=10pt,
    enlarge y limits=0.12,
    xlabel={Percent},
    ylabel={Response},
    font=\footnotesize,
    ytick pos=left,
    symbolic y coords={5,4,3,2,1},
    ytick=data,
    nodes near coords,
    every node near coord/.append style={anchor=west},
]
\addplot[
    draw=none,
    fill=blue!70!black
] coordinates {
    (18.6,5) [18.6\%]
    (44.7,4) [44.7\%]
    (23.6,3) [23.6\%]
    (7.3,2)  [7.3\%]
    (5.7,1)  [5.7\%]
};
\end{axis}
\end{tikzpicture}

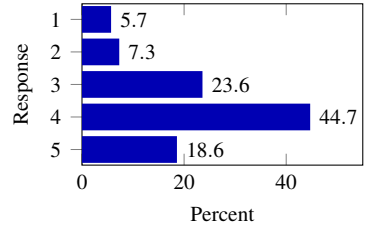
\captionof{figure}{Student view on integration of AI literacy in academic curriculum on scale of 1-5, where 1 represent strongly disagree and 5 represent strongly agree.}
\label{Q13}
\end{minipage}

Figure \ref{Q12} indicates strong student support for the formal inclusion of AI-related skills within programme Knowledge, Skills, and Behaviours (KSBs). A total of 63.8\% of respondents express agreement, including 47.0\% who agree and 16.8\% who strongly agree. In comparison, 24.4\% of students remain neutral, while only a small proportion report disagreement (6.0\% disagree and 5.8\% strongly disagree). This clear student mandate for formalizing AI skills in KSBs reflects underlying employability anxiety ~\cite{liang2025systematic, deep2025evaluating}. Students increasingly view AI literacy as an essential professional competence rather than a short-term academic aid. Their demand for formal inclusion signals a desire for universities to recognise and validate the time spent developing these skills. By seeking this recognition, students aim to legitimise learning that currently occurs outside the formal curriculum and to ensure that their degrees communicate relevant technical competencies to employers. However, this expectation presents a challenge for universities, as formal inclusion requires reliable assessment practices, and many institutions currently lack clear frameworks or expertise to evaluate AI-related practices such as interactive use of chat-based tools.\\
\textbf{Q13: Should AI literacy be formally integrated into your curriculum?}

Figure \ref{Q13} shows strong support for the formal integration of AI literacy into the academic curriculum. A total of 63.3\% of respondents express agreement, including 44.7\% who agree and 18.6\% who strongly agree. In contrast, 23.6\% of students select a neutral position, while a relatively small proportion report disagreement (7.3\% disagree and 5.7\% strongly disagree). Similar to the mandate for inclusion in Knowledge, Skills, and Behaviours, this level of agreement reflects a broader fear of academic and professional obsolescence. Students increasingly perceive degrees without AI training as outdated and believe this places them at a disadvantage compared to graduates from institutions that have already adapted. Their call for formal integration ~\cite{liang2025systematic, luo2025design} also reveals a mismatch in expectations. Students operate within the fast pace of technological change, while universities function within slower cycles of accreditation and curriculum approval. This difference in timelines creates tension between student demand and institutional capacity for reform. In addition, students may not fully recognise that formal integration requires formal assessment, which could result in them being evaluated on skills they have largely developed independently.\\

\noindent
\begin{minipage}[t]{0.49\linewidth}
\centering
\begin{tikzpicture}
\begin{axis}[
    xbar,
    width=\linewidth,
    height=0.7\linewidth,
    xmin=0,
    xmax=60,
    bar width=10pt,
    enlarge y limits=0.12,
    xlabel={Percent},
    ylabel={Response},
    font=\footnotesize,
    ytick pos=left,
    symbolic y coords={5,4,3,2,1},
    ytick=data,
    nodes near coords,
    every node near coord/.append style={anchor=west},
]
\addplot[
    draw=none,
    fill=blue!70!black
] coordinates {
    (17.7,5) [17.7\%]
    (46.7,4) [46.7\%]
    (27.1,3) [27.1\%]
    (5.3,2)  [5.3\%]
    (3.2,1)  [3.2\%]
};
\end{axis}
\end{tikzpicture}

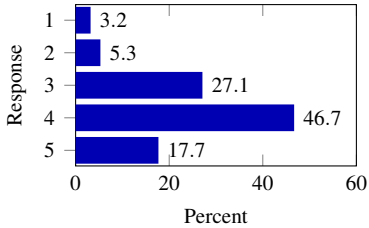
\captionof{figure}{Student satisfaction levels with university communication and transparent on GenAI integration in courses on a scale of 1-5, where 1 represent very dissatisfied and 5 represent very satisfied.}
\label{Q14}
\end{minipage}\hfill
\begin{minipage}[t]{0.49\linewidth}
\centering
\begin{tikzpicture}
\begin{axis}[
    xbar,
    width=\linewidth,
    height=0.7\linewidth,
    xmin=0,
    xmax=48,
    bar width=10pt,
    enlarge y limits=0.12,
    xlabel={Percent},
    ylabel={Response},
    font=\footnotesize,
    ytick pos=left,
    symbolic y coords={5,4,3,2,1},
    ytick=data,
    nodes near coords,
    every node near coord/.append style={anchor=west},
]
\addplot[
    draw=none,
    fill=blue!70!black
] coordinates {
    (10.6,5) [10.6\%]
    (38.8,4) [38.8\%]
    (35.1,3) [35.1\%]
    (11.5,2) [11.5\%]
    (4.1,1)  [4.1\%]
};
\end{axis}
\end{tikzpicture}

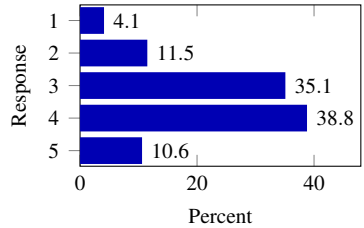
\captionof{figure}{Student view regarding lecturers encouragements for AI use in curriculum.}
\label{Q15}
\end{minipage}

\textbf{Q14: How satisfied are you with the university's communication?}

Figure \ref{Q14} shows generally high levels of satisfaction with institutional communication and transparency regarding AI use. A total of 64.4\% of respondents report being satisfied, including 46.7\% who are satisfied and 17.7\% who are very satisfied. In comparison, 27.1\% of students select a neutral response, while only a small minority express dissatisfaction (5.3\% dissatisfied and 3.2\% very dissatisfied). This high level of satisfaction reflects what can be described as a transparency placebo ~\cite{nguyen2025use, kim2025examining}. Students feel reassured because clear communication reduces uncertainty around acceptable and unacceptable uses of AI, particularly in relation to academic misconduct. However, satisfaction with communication should not be interpreted as satisfaction with learning support. While the university has effectively reduced ambiguity anxiety by clarifying rules and boundaries, this outcome represents a defensive achievement rather than an educational one. It demonstrates success in policy communication and risk management, but provides limited insight into the quality or availability of pedagogical guidance that would enable students to use AI as a meaningful learning tool.\\
\textbf{Q15. Do lecturers actively encourage the exploration of AI?}

Figure \ref{Q15} shows a divided pattern in students’ perceptions of lecturer encouragement to explore and use AI tools within the curriculum. A total of 49.4\% of respondents report encouragement, including 38.8\% who agree and 10.6\% who strongly agree. At the same time, 35.1\% of students select a neutral position, while 15.6\% express disagreement (11.5\% disagree and 4.1\% strongly disagree). This uneven distribution points to an inconsistency crisis in institutional practice. The high level of neutrality suggests that a significant proportion of lecturers are avoiding engagement with AI-related discussions altogether. In the current educational context, such neutrality effectively functions as discouragement, as it leaves students without clear guidance. As a result, students may rely on informal or hidden forms of AI use that fall outside transparent academic practice. This inconsistency undermines institutional equity, since students in different courses or sections may receive conflicting messages about acceptable AI use. Such variation increases student anxiety and creates conditions for future disputes over assessment fairness, particularly when access to and guidance on learning tools are not applied consistently across the institution.\\

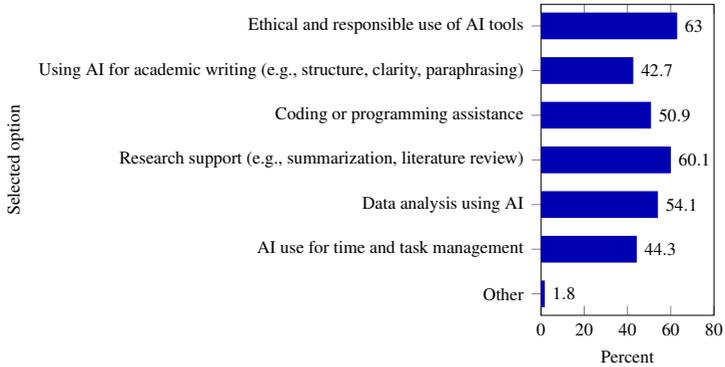
\begin{figure}[t]
\centering
\resizebox{0.9\linewidth}{!}{\begin{tikzpicture}
\begin{axis}[
    xbar,
    width=0.40\linewidth,
    height=0.6\linewidth,
    xmin=0,
    xmax=80,
    bar width=12pt,
    enlarge y limits=0.08,
    xlabel={Percent},
    ylabel={Selected option},
    font=\footnotesize,
    ytick pos=left,
    symbolic y coords={
        Other,
        {AI use for time and task management},
        {Data analysis using AI},
        {Research support (e.g., summarization, literature review)},
        {Coding or programming assistance},
        {Using AI for academic writing (e.g., structure, clarity, paraphrasing)},
        {Ethical and responsible use of AI tools}
    },
    ytick=data,
    nodes near coords,
    every node near coord/.append style={anchor=west},
]
\addplot[
    draw=none,
    fill=blue!70!black
] coordinates {
    (1.8,Other) [1.8\%]
    (44.3,{AI use for time and task management}) [44.3\%]
    (54.1,{Data analysis using AI}) [54.1\%]
    (60.1,{Research support (e.g., summarization, literature review)}) [60.1\%]
    (50.9,{Coding or programming assistance}) [50.9\%]
    (42.7,{Using AI for academic writing (e.g., structure, clarity, paraphrasing)}) [42.7\%]
    (63.0,{Ethical and responsible use of AI tools}) [63.0\%]
};
\end{axis}
\end{tikzpicture}}
\caption{Student preferences for Generative AI training topics.}
\label{Q16}
\end{figure}

\noindent
\newcommand{\QSixteenHeight}{0.22\textheight}
\begin{minipage}[t]{0.50\linewidth}
\centering
\fbox{\includegraphics[width=0.95\linewidth,height=\QSixteenHeight]{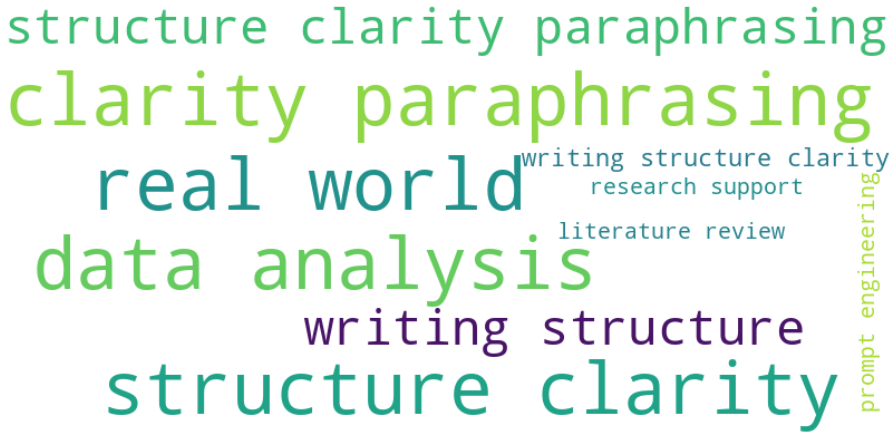}}
\vspace{4pt}
\captionsetup{hypcap=false}
\captionof{figure}{Word cloud showing student preferences for Generative AI training topics.}
\label{Q16.b}
\end{minipage}\hfill
\begin{minipage}[t]{0.49\linewidth}
\centering
\resizebox{\linewidth}{\QSixteenHeight}{%
\begin{tikzpicture}
\begin{axis}[
    xbar,
    width=\linewidth,
    height=0.8\linewidth,
    xmin=0,
    xmax=55,
    bar width=10pt,
    enlarge y limits=0.15,
    xlabel={Percent},
    ylabel={Response},
    font=\footnotesize,
    ytick pos=left,
    symbolic y coords={5,4,3,2,1},
    ytick=data,
    nodes near coords,
    every node near coord/.append style={anchor=west},
]
\addplot[
    draw=none,
    fill=blue!70!black
] coordinates {
    (19.4,5) [19.4\%]
    (45.6,4) [45.6\%]
    (22.1,3) [22.1\%]
    (7.6,2)  [7.6\%]
    (5.3,1)  [5.3\%]
};
\end{axis}
\end{tikzpicture}
}
\vspace{4pt}
\captionsetup{hypcap=false}

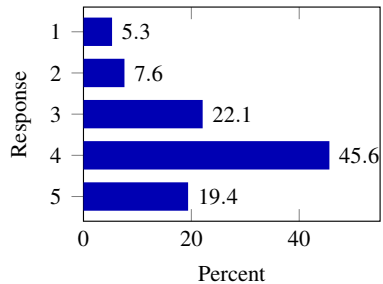
\captionof{figure}{Student expectations for instructor-led guidance on AI usage in assessments on a scale of 1–5, where 1 represents strongly disagree and 5 represents strongly agree.}
\label{Q18}
\end{minipage}

\textbf{Q16 \& Q16 (b): Which types of AI training would be most beneficial?}\\

Figure \ref{Q16}, which reports multiple responses, shows that the ethical and responsible use of AI tools is the most frequently selected training priority, identified by 63.8\% of respondents. This is followed by interest in research support at 60.1\%, data analysis using AI at 54.1\%, and coding or programming assistance at 50.9\%. Smaller but still substantial proportions of students express interest in using AI for academic writing (42.7\%) and for time and task management (44.3\%). The prominence of ethical and responsible use as the top priority suggests a strong safety first orientation among students. Rather than reflecting a demand for abstract ethical discussion, this preference appears to signal a need for clear guidance on acceptable use and on how to avoid academic misconduct. Students are seeking precise boundaries that allow them to use AI tools with confidence. The word cloud in Fig. \ref{Q16.b} reinforces this interpretation by emphasising terms such as structure, clarity, and paraphrasing, which point to practical concerns about improving written work. This pattern indicates that students view AI primarily as a tool for refining and editing academic output. At the same time, strong demand for research support and data analysis suggests a desire to reduce routine technical tasks, enabling students to concentrate on higher-level analysis and synthesis.\\
\textbf{Q18: I expect course tutors to provide clear guidance?}

Figure \ref{Q18} shows strong student expectations for explicit guidance on the use of AI in assessments. A total of 65.2\% of respondents express agreement, including 45.6\% who agree and 19.4\% who strongly agree. In comparison, 22.1\% of students select a neutral response, while only a small minority report disagreement (7.6\% disagree and 5.3\% strongly disagree). This high level of agreement indicates that students perceive guidance on AI use as a basic academic entitlement rather than optional support. Students show low tolerance for ambiguity and increasingly treat assessment rubrics as formal agreements that define acceptable practice. When instructions on AI use are not clearly stated, students report feeling uncertain and exposed to risk. This pattern reflects a broader flight from ambiguity ~\cite{pudasaini2024survey, kim2025examining}, which places pressure on universities to provide highly detailed and standardised assessment guidance. In the absence of explicit instructions, students may interpret later allegations of misconduct as unfair, using the lack of prior guidance as a mitigating factor in academic integrity processes.\\
\textbf{Q19: In which parts of the curriculum would you like integration the most?} 

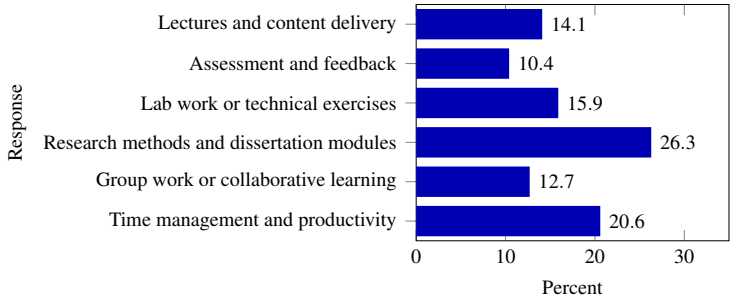
\begin{figure}[t]
\centering
\resizebox{0.9\linewidth}{!}{\begin{tikzpicture}
\begin{axis}[
    xbar,
    width=0.55\linewidth,
    height=0.45\linewidth,
    xmin=0,
    xmax=35,
    bar width=12pt,
    enlarge y limits=0.1,
    xlabel={Percent},
    ylabel={Response},
    font=\footnotesize,
    ytick pos=left,
    symbolic y coords={
        {Time management and productivity},
        {Group work or collaborative learning},
        {Research methods and dissertation modules},
        {Lab work or technical exercises},
        {Assessment and feedback},
        {Lectures and content delivery}
    },
    ytick=data,
    nodes near coords,
    every node near coord/.append style={anchor=west},
]
\addplot[
    draw=none,
    fill=blue!70!black
] coordinates {
    (20.6,{Time management and productivity}) [20.6\%]
    (12.7,{Group work or collaborative learning}) [12.7\%]
    (26.3,{Research methods and dissertation modules}) [26.3\%]
    (15.9,{Lab work or technical exercises}) [15.9\%]
    (10.4,{Assessment and feedback}) [10.4\%]
    (14.1,{Lectures and content delivery}) [14.1\%]
};
\end{axis}
\end{tikzpicture}}
\caption{Student-identified priorities for the embedding of AI competencies in curriculum.}
\label{f:Q19}
\end{figure}

Figure \ref{f:Q19} shows that students express the strongest preference for integrating AI tools within research methods and dissertation modules, selected by 26.3\% of respondents. This is followed by time management and productivity at 20.6\% and lab work or technical exercises at 15.9\%. Lower levels of preference are reported for lectures and content delivery (14.1\%), group work or collaborative learning (12.7\%), and especially assessment and feedback (10.4\%). The strong preference for AI integration in research methods and dissertation modules ~\cite{luo2025design, liang2025systematic} reflects students’ desire for support in managing the most demanding component of their degree. The dissertation requires sustained independent work, complex reading, and methodological planning, and students appear to view AI as a tool to reduce the workload associated with these processes. This trend presents a significant challenge for universities, as the dissertation has traditionally served as evidence of independent research capability. Integrating AI into this space raises questions about how academic independence and originality should be defined. At the same time, the relatively low interest in AI use for assessment and feedback suggests that students prefer AI as a support tool rather than an evaluative authority, indicating limited trust in automated judgment within academic assessment.\\
\textbf{Q20 \& Q20(b): In what ways might GenAI negatively affect your experience?}

\begin{figure}[t]
\centering
\resizebox{0.9\linewidth}{!}{\begin{tikzpicture}
\begin{axis}[
    xbar,
    width=0.50\linewidth,
    height=0.55\linewidth,
    xmin=0,
    xmax=75,
    bar width=10pt,
    enlarge y limits=0.08,
    xlabel={Percent},
    ylabel={Selected Option},
    font=\footnotesize,
    ytick pos=left,
    symbolic y coords={
        Other,
        {I do not believe AI tools negatively affect my academic experience},
        {Insufficient guidance or training from lecturers or university},
        {Data privacy and ethical concerns},
        {Low-quality or inaccurate AI outputs},
        {Unclear rules about plagiarism or academic misconduct},
        {Risk of over-reliance and reduced critical thinking}
    },
    ytick=data,
    nodes near coords,
    every node near coord/.append style={anchor=west},
]
\addplot[
    draw=none,
    fill=blue!70!black
] coordinates {
    (1.1,Other) [1.1\%]
    (14.7,{I do not believe AI tools negatively affect my academic experience}) [14.7\%]
    (33.7,{Insufficient guidance or training from lecturers or university}) [33.7\%]
    (53.6,{Data privacy and ethical concerns}) [53.6\%]
    (43.0,{Low-quality or inaccurate AI outputs}) [43.0\%]
    (43.9,{Unclear rules about plagiarism or academic misconduct}) [43.9\%]
    (56.1,{Risk of over-reliance and reduced critical thinking}) [56.1\%]
};
\end{axis}
\end{tikzpicture}}
\caption{Student concerns regarding the negative impact of Generative AI on the academic experience.}
\label{Q20}
\end{figure}
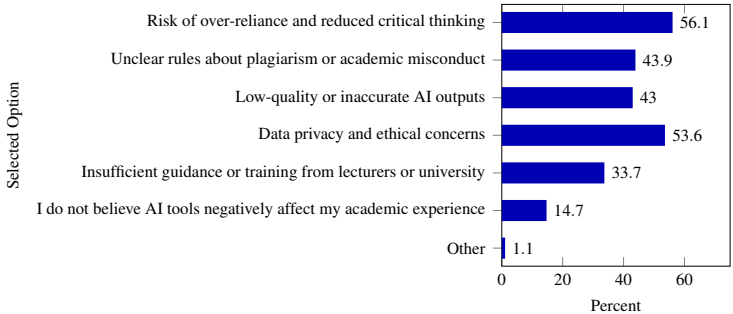

% height control
\renewcommand{\QSixteenHeight}{0.22\textheight}
\begin{minipage}[t]{0.49\linewidth}
\centering
\fbox{\includegraphics[width=0.95\linewidth,height=\QSixteenHeight]{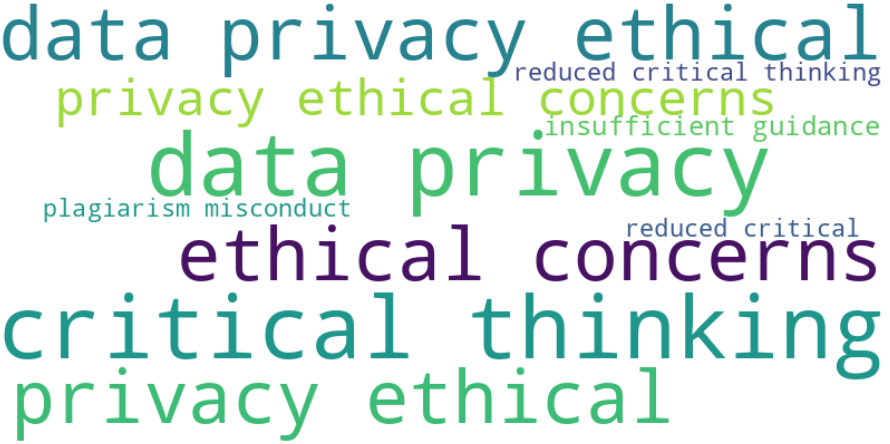}}
\vspace{4pt}
\captionsetup{hypcap=false}
\captionof{figure}{Word cloud showing student preferences for Generative AI training topics.}
\label{Q20.b}
\end{minipage}\hfill
\begin{minipage}[t]{0.49\linewidth}
\centering
\resizebox{\linewidth}{\QSixteenHeight}{%
\begin{tikzpicture}
\begin{axis}[
    xbar,
    width=\linewidth,
    height=0.7\linewidth,
    xmin=0,
    xmax=50,
    bar width=10pt,
    enlarge y limits=0.12,
    xlabel={Percent},
    ylabel={Response},
    font=\footnotesize,
    ytick pos=left,
    symbolic y coords={5,4,3,2,1},
    ytick=data,
    nodes near coords,
    every node near coord/.append style={anchor=west},
]
\addplot[
    draw=none,
    fill=blue!70!black
] coordinates {
    (5.8,5)  [5.8\%]
    (19.2,4) [19.2\%]
    (38.4,3) [38.4\%]
    (22.7,2) [22.7\%]
    (13.9,1) [13.9\%]
};
\end{axis}
\end{tikzpicture}
}
\vspace{4pt}
\captionsetup{hypcap=false}

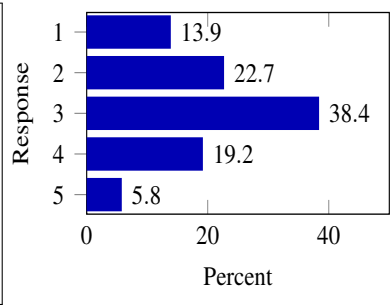
\captionof{figure}{Student expectations for instructor-led guidance on AI usage in assessments on a scale of 1–5, where 1 represents strongly disagree and 5 represents strongly agree.}
\label{Q22}
\end{minipage}

Figure \ref{Q20} shows that students report several major concerns about the potential negative impact of generative AI on their academic experience. Reporting multiple responses to this question, the most frequently cited concern is the risk of over-reliance and reduced critical thinking, selected by 56.1\% of respondents. This is followed by data privacy and ethical concerns at 53.6\%. Significant proportions of students also identify unclear rules related to plagiarism or academic misconduct (43.9\%) and concerns about low-quality or inaccurate AI outputs (43.0\%). A smaller but notable group highlights insufficient guidance or training from lecturers or the university (33.7\%). Only 14.7\% of respondents indicate that they do not believe AI tools have a negative effect on their academic experience. These findings suggest a high level of student awareness of what can be described as cognitive atrophy, with reduced critical thinking emerging as the primary concern. Students appear to recognise that while AI increases efficiency, it can also discourage deeper engagement unless limits are in place. Qualitative responses in Fig. \ref{Q20.b} support this interpretation, as reduced critical thinking is repeatedly mentioned alongside privacy and ethical concerns. This pattern indicates that students are not rejecting AI use, but are instead seeking institutional structures, such as assessment designs that require active reasoning, to counterbalance over-reliance. In addition, strong concern about data privacy points to a critical awareness of the risks associated with commercial AI platforms, challenging assumptions that students are unconcerned about how their data are used.\\
\textbf{Q22: Has reliance on AI reduced opportunities to demonstrate ability?}

Figure \ref{Q22} shows that a majority of students believe increased reliance on AI tools reduces opportunities to demonstrate their own academic abilities. In total, 63.4\% of respondents report a moderate to extreme impact, including 38.4\% who select ‘moderately’, 19.2\% ‘very much’, and 5.8\% ‘extremely’. In comparison, 22.7\% report only a slight impact, while 13.9\% indicate no impact at all. These responses suggest that students perceive AI use as flattening differences in academic performance. This process, described as a homogenization of merit ~\cite{kofinas2025impact, pudasaini2024survey}, appears to raise the baseline level of work while reducing the visibility of exceptional performance. High-achieving students may feel that their originality and personal voice are less distinguishable when polished and well-structured outputs can be easily produced with AI assistance. As a result, the signalling value of grades may be weakened if average-quality work becomes easier to generate. This finding indicates that current assessment practices may unintentionally favour AI-like characteristics such as clarity, structure, and grammatical accuracy over less refined but original human thinking, contributing to a sense of credential inflation in which high grades feel less strongly earned.\\
\textbf{Q23: How well prepared are your lecturers?}

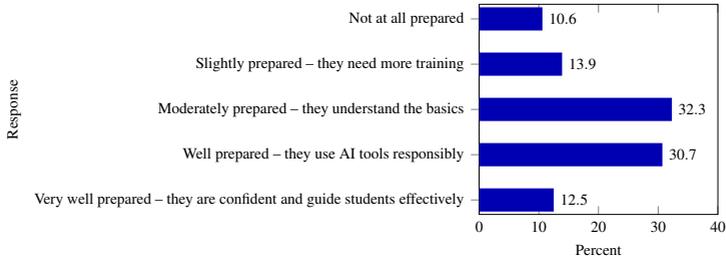
\begin{figure}[t]
\centering
\resizebox{0.9\linewidth}{!}{\begin{tikzpicture}
\begin{axis}[
    xbar,
    width=0.55\linewidth,
    height=0.5\linewidth,
    xmin=0,
    xmax=40,
    bar width=12pt,
    enlarge y limits=0.08,
    xlabel={Percent},
    ylabel={Response},
    font=\footnotesize,
    ytick pos=left,
    symbolic y coords={
        {Very well prepared -- they are confident and guide students effectively},
        {Well prepared -- they use AI tools responsibly},
        {Moderately prepared -- they understand the basics},
        {Slightly prepared -- they need more training},
        {Not at all prepared}
    },
    ytick=data,
    nodes near coords,
    every node near coord/.append style={anchor=west},
]
\addplot[
    draw=none,
    fill=blue!70!black
] coordinates {
    (12.5,{Very well prepared -- they are confident and guide students effectively}) [12.5\%]
    (30.7,{Well prepared -- they use AI tools responsibly}) [30.7\%]
    (32.3,{Moderately prepared -- they understand the basics}) [32.3\%]
    (13.9,{Slightly prepared -- they need more training}) [13.9\%]
    (10.6,{Not at all prepared}) [10.6\%]
};
\end{axis}
\end{tikzpicture}}
\caption{Student assessment of lecturer preparedness regarding GenAI for teaching and assessments.}
\label{f:Q23}
\end{figure}

Figure \ref{f:Q23} shows mixed student perceptions of lecturer preparedness to use AI tools effectively in teaching and assessment. A total of 63.0\% of respondents view their lecturers as at least moderately prepared, including 32.3\% who select moderately prepared and 30.7\% who report well prepared. In contrast, 24.5\% of students perceive lower levels of readiness, with 13.9\% describing lecturers as slightly prepared and in need of further training, and 10.6\% viewing them as not prepared at all. Only 12.5\% of respondents consider lecturers to be very well prepared to confidently guide students in AI use. These findings highlight a clear competence gap ~\cite{mah2024artificial, kim2025examining}, as nearly one quarter of students question their lecturers’ readiness in this area. In higher education, academic authority is closely linked to perceived expertise, and doubts about technological competence can weaken this authority. The large proportion of moderate ratings suggests that students believe lecturers are managing the transition to AI rather than actively leading it. This perception may be influenced by a digital native bias, where students assess preparedness based on visible technical fluency rather than deeper pedagogical understanding, potentially confusing difficulties with interfaces for a lack of insight into the broader educational implications of AI.\\

\renewcommand{\QSixteenHeight}{0.22\textheight}

\noindent
\begin{minipage}[t]{0.49\linewidth}
\centering
\resizebox{\linewidth}{\QSixteenHeight}{%
  \begin{tikzpicture}
\begin{axis}[
    xbar,
    width=\linewidth,
    height=0.8\linewidth,
    xmin=0,
    xmax=45,
    bar width=10pt,
    enlarge y limits=0.12,
    xlabel={Percent},
    ylabel={Response},
    font=\footnotesize,
    ytick pos=left,
    symbolic y coords={5,4,3,2,1},
    ytick=data,
    nodes near coords,
    every node near coord/.append style={anchor=west},
]
\addplot[
    draw=none,
    fill=blue!70!black
] coordinates {
    (16.8,5) [16.8\%]
    (29.3,4) [29.3\%]
    (35.0,3) [35.0\%]
    (10.8,2) [10.8\%]
    (8.1,1)  [8.1\%]
};
\end{axis}
\end{tikzpicture}%
}
\vspace{4pt}
\captionsetup{hypcap=false}

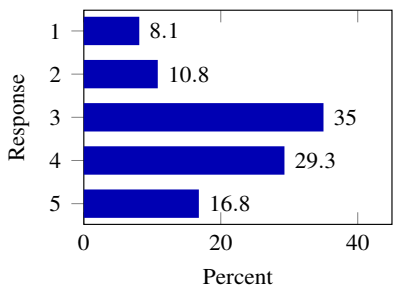
\captionof{figure}{Student attitudes toward the mandatory disclosure of AI-generated content in assessments on a scale of 1–5, where 1 represents strongly disagree and 5 represents strongly agree.}
\label{Q24}
\end{minipage}\hfill
\begin{minipage}[t]{0.49\linewidth}
\centering
\resizebox{\linewidth}{\QSixteenHeight}{%
  \begin{tikzpicture}
\begin{axis}[
    xbar,
    width=\linewidth,
    height=0.8\linewidth,
    xmin=0,
    xmax=45,
    bar width=10pt,
    enlarge y limits=0.12,
    xlabel={Percent},
    ylabel={Response},
    font=\footnotesize,
    ytick pos=left,
    symbolic y coords={5,4,3,2,1},
    ytick=data,
    nodes near coords,
    every node near coord/.append style={anchor=west},
]
\addplot[
    draw=none,
    fill=blue!70!black
] coordinates {
    (20.5,5) [20.5\%]
    (36.2,4) [36.2\%]
    (31.8,3) [31.8\%]
    (7.4,2)  [7.4\%]
    (4.1,1)  [4.1\%]
};
\end{axis}
\end{tikzpicture}%
}
\vspace{4pt}
\captionsetup{hypcap=false}

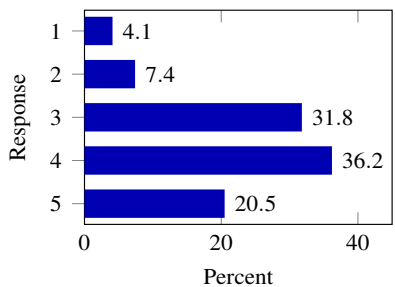
\captionof{figure}{Student expectation on the institutional responsibility to provide equitable access to GenAI tools on a scale of 1–5, where 1 represents strongly disagree and 5 represents strongly agree.}
\label{Q25}
\end{minipage}

\textbf{Q24: Should students be required to declare AI-generated content?}

Figure \ref{Q24} shows mixed but generally cautious support for requiring students to declare AI-generated content in academic submissions. A total of 46.1\% of respondents express agreement, including 29.3\% who agree and 16.8\% who strongly agree. At the same time, a large proportion of students select a neutral response (35.0\%), while a smaller minority report disagreement (10.8\% disagree and 8.1\% strongly disagree). Support for mandatory declaration reflects what can be described as an honesty box approach ~\cite{pudasaini2024survey}. Students appear to view declaration as a protective administrative measure that allows them to legitimise their AI use and reduce anxiety about potential misconduct. However, this approach raises an enforceability paradox. Students who use AI in limited or permitted ways are more likely to declare their use, while those who rely on AI in inappropriate ways may be least willing to self-report. As a result, the policy risks capturing information mainly from compliant students while failing to address more serious misuse. The large neutral group further suggests that many students are indifferent to the mechanism of declaration itself and are primarily concerned with whether the process provides a sense of security rather than ensuring substantive academic integrity.\\
\textbf{Q25: Should universities ensure equal access to GenAI tools?}

Figure \ref{Q25}, the data reveals a significant consensus regarding the institutional responsibility to bridge the digital divide in artificial intelligence. A clear majority of students support the mandate for equal access, with 36.2\% agreeing and 20.5\% strongly agreeing, resulting in a combined positive sentiment of 56.7\%. In contrast, opposition is minimal, with only 11.5\% of respondents expressing disagreement (7.4\% disagreeing and 4.1\% strongly disagreeing). A notable 31.8\% of the cohort remained neutral, a trend that may reflect "Privilege Blindness" among those who already possess the financial means to access premium tools.
This majority agreement underscores the "Algorithmic Paywall" as discussed  crisis, as discussed in ~\cite{mah2024artificial, kim2025examining}. Students recognize that a "Two-Tier" system has emerged, where wealthier students access superior models while others must rely on inferior free versions. By demanding equal access, students are arguing that AI is no longer a luxury good but has become essential academic infrastructure. For the university, this presents a significant budgetary challenge, complying with this mandate requires massive investment, yet failing to do so effectively endorses a "pay-to-win" system where financial capital directly translates into academic advantage ~\cite{mah2024artificial, kim2025examining}.\\
\textbf{Q26 \& Q26(b): What would improve awareness of AI policies?}

\renewcommand{\QFiveLeftBlockWidth}{0.66\linewidth}
\renewcommand{\QFiveRightBlockWidth}{0.34\linewidth}

\renewcommand{\QFiveLeftImgHeight}{0.32\textheight}
\renewcommand{\QFiveRightImgHeight}{0.19\textheight}

\noindent
\begin{minipage}[t]{\QFiveLeftBlockWidth}
\centering
\resizebox{\linewidth}{\QFiveLeftImgHeight}{%
  \begin{tikzpicture}
\begin{axis}[
    xbar,
    width=0.45\linewidth,
    height=0.7\linewidth,
    xmin=0,
    xmax=70,
    bar width=10pt,
    enlarge y limits=0.08,
    xlabel={Percent},
    ylabel={Selected option},
    font=\tiny,
    ytick pos=left,
    symbolic y coords={
        Other,
        {Integrating AI guideline directly into coursework materials},
        {Regular workshops or webinars about AI guidelines},
        {Mandatory quizzes on AI guidelines at the beginning of the term},
        {Dedicated academic staff or help desk available to address queries},
        {Frequently Asked Questions sections on university portals},
        {Clearly visible links on course homepages}
    },
    ytick=data,
    nodes near coords,
    every node near coord/.append style={anchor=west},
]
\addplot[
    draw=none,
    fill=blue!70!black
] coordinates {
    (0.7,Other) [0.7\%]
    (36.9,{Integrating AI guideline directly into coursework materials}) [36.9\%]
    (50.2,{Regular workshops or webinars about AI guidelines}) [50.2\%]
    (45.6,{Mandatory quizzes on AI guidelines at the beginning of the term}) [45.6\%]
    (37.6,{Dedicated academic staff or help desk available to address queries}) [37.6\%]
    (33.7,{Frequently Asked Questions sections on university portals}) [33.7\%]
    (33.0,{Clearly visible links on course homepages}) [33.0\%]
};
\end{axis}
\end{tikzpicture}%
}
\vspace{4pt}
\captionsetup{hypcap=false}
\captionof{figure}{Student preferred methods for improving student awareness of AI guidelines.}
\label{Q26}
\end{minipage}\hfill
\begin{minipage}[t]{\QFiveRightBlockWidth}
\centering
\raisebox{0.09\textheight}{%
\fbox{\includegraphics[width=0.9\linewidth,height=\QFiveRightImgHeight]{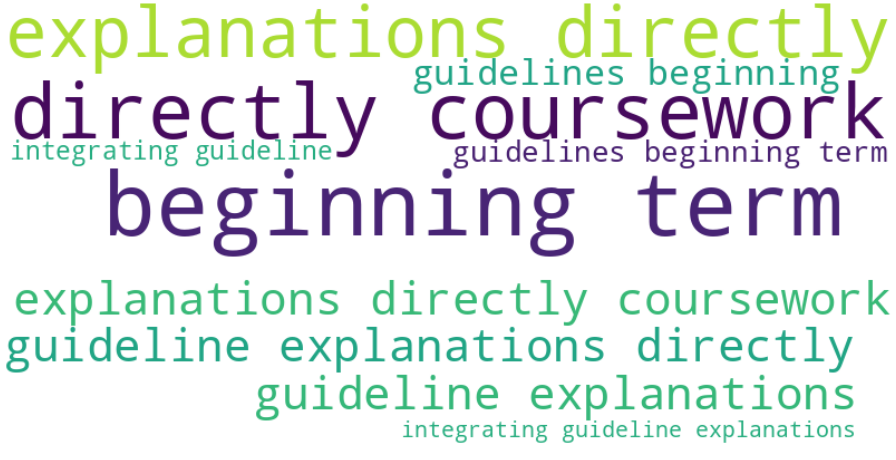}}
}
\vspace{6pt}
\captionsetup{hypcap=false}
\captionof{figure}{Word cloud for preferred methods for improving student awareness of AI guidelines.}
\label{Q26.b}
\end{minipage}

Figure \ref{Q26} shows broad student support for equitable access to generative AI tools. A total of 56.7\% of respondents express agreement, including 36.2\% who agree and 20.5\% who strongly agree. In addition, 31.8\% of students select a neutral position, while only a small minority report disagreement (7.4\% disagree and 4.1\% strongly disagree). These responses indicate that students are moving beyond passive forms of communication, such as policy documents or email notices, and instead favour active and timely forms of learning. Preferences for mandatory quizzes and integration into coursework suggest a desire for structured confirmation of acceptable practice. Students appear to seek formal assurance that they understand institutional rules, which they believe will protect them from future accusations of academic misconduct. The word cloud shown in Fig. \ref{Q26.b}, which highlights terms such as explanations, directly, and coursework, supports this interpretation by showing that students want clear, applied guidance rather than general statements. Overall, the findings suggest that students value practical, embedded instruction on AI use that provides both clarity and a sense of procedural security.
\section{Discussion}
%\section{Pedagogical Importance of %the Study}
This study makes a clear pedagogical contribution by showing that student engagement with Generative AI in higher education is shaped less by basic awareness and more by guidance, confidence, and academic framing. Although students report high familiarity with GenAI, actual academic use is lower and confidence in effective use remains moderate. This gap suggests that technical fluency does not automatically produce critical or responsible use. The findings therefore support treating AI literacy as a core area of learning rather than a peripheral digital skill ~\cite{Kelly2023Generative, Kasneci2023ChatGPT}.

The reported uses of GenAI concept clarification, brainstorming, and summarising indicate that many students use AI as a cognitive support tool rather than a direct substitute for their own work. Pedagogically, this matters because it shifts the focus from prohibition to structured integration. The key issue is not only whether students use AI, but how institutions help them use it for judgment, verification, and deeper understanding.

A second contribution is the evidence of mismatch between student demand and institutional provision. Strong support for AI literacy in the curriculum, high interest in training, and expectations for tutor guidance suggest that students see GenAI as part of mainstream learning, not an external add-on. The findings also highlight the need for explicit structures for ethical practice, evaluative thinking, and appropriate academic use ~\cite{Yusuf2024GenerativeAI, Chan2023StudentsVoices}.

The study is also relevant to assessment design. Student concerns about over-reliance, reduced critical thinking, privacy risks, and incorrect outputs indicate that the challenge is not only AI use itself, but how to assess independent reasoning in AI-enabled contexts. These results align with calls for assessment approaches that prioritise process, justification, and transparency over product-only evaluation ~\cite{Xia2024ScopingReview}.

Finally, the findings show that pedagogical change depends on institutional readiness as well as student readiness. Mixed perceptions of lecturer preparedness and concerns about unequal tool access suggest that effective implementation requires both staff development and equity-focused infrastructure. Overall, the study identifies practical gaps that institutions must address if GenAI is to be implemented as an educationally meaningful practice rather than a purely administrative control problem.

\section{Conclusion}
The integration of GenAI into higher education is no longer a prospective shift but a present reality. This study of 436 Computer Science postgraduates reveals that students increasingly view GenAI as an essential academic infrastructure, comparable to internet access. However, a significant gap exists between student awareness and institutional support. While the university has successfully communicated its restrictive policies, leading to high levels of \textit{awareness} it has been less effective in providing the practical, licensed resources and pedagogical training students actively seek.

The data highlights several critical areas for institutional reform:
\begin{itemize}

\item From restriction to integration: Students do not view AI as a replacement for educators but as a mentor-led \textit{human-AI partnership}. There is a clear mandate for AI literacy to be embedded directly into curricula and formalized in KSBs to ensure graduates are workplace-ready.

\item Addressing the \textit{confidence-competence paradox}: Institutions must help students move beyond \textit{usage familiarity} toward \textit{critical familiarity} ensuring they can identify hallucinations, bias, and inaccuracies.

\item Ensuring equity: To prevent a \textit{two-tier} education system, universities must mitigate the \textit{Algorithmic Paywall} by providing equitable access to premium GenAI models for all students, regardless of financial standing.
\end{itemize}

Ultimately, the goal of higher education in the AI era should not be the detection of AI use, but the cultivation of \textit{higher-order} human skills-critical thinking, ethical judgment, and originality-that GenAI cannot replicate. By moving from a \textit{policing} to a \textit{pedagogical} approach, institutions can empower students to become informed, ethical, and innovative co-creators in an AI-saturated world.

\subsection*{Limitations of the Study}
\begin{itemize}
\item The study is based on a single institutional setting, targeting postgraduate computer science students, which may affect its generalizability to other disciplines.
\item The study is based on self-reported data, and participants' responses could be influenced by various biases, including social desirability bias or uncertainty about institutional norms related to AI use.
\item The study is descriptive in nature, establishing patterns related to familiarity, usage, and perceptions of Generative AI, but not testing causal relationships between variables.
\end{itemize}

\subsection*{Future Research Directions}
\begin{itemize}
\item Future studies could be conducted in cross-disciplinary and cross-institutional settings to establish differences in AI usage and perceptions across varied academic contexts.
\item Mixed-methods studies, including interviews, observations, or analyses of learning analytics, could offer deeper insights into students' actual AI usage in their academic practices.
\item Longitudinal studies are needed as Generative AI becomes increasingly integrated into higher education, affecting students' capabilities, perceptions, and pedagogical approaches over time.
\end{itemize}

%%%%%%%%%%%%%%%%%%%%%%%%%%%%%%%%%%%%
% REFERENCES
%%%%%%%%%%%%%%%%%%%%%%%%%%%%%%%%%%%%
\bibliographystyle{dcu-rbfin}
\bibliography{Reference_list}

@article{AlEmran2018Investigating,
  author = {Al-Emran, Mahmoud and Elsherif, Husam M. and Shaalan, Khaled},
  title = {Investigating attitudes towards the use of mobile learning in higher education},
  journal = {Computers in Human Behavior},
  volume = {56},
  pages = {93--102},
  year = {2018},
  doi = {10.1016/j.chb.2015.11.033}
}

@article{Ardito2024GenerativeAI,
  author = {Ardito, Carmina G.},
  title = {Generative AI detection in higher education assessments},
  journal = {New Directions for Teaching and Learning},
  volume = {2024},
  issue = {1},
  pages = {1--18},
  year = {2024},
  doi = {10.1002/tl.20624}
}

@article{Arowosegbe2024Perception,
  author = {Arowosegbe, Anuoluwapo and Alqahtani, Jassem S. and Oyelade, Toluwanimi},
  title = {Perception of generative AI use in UK higher education},
  journal = {Frontiers in Education},
  volume = {9},
  pages = {1463208},
  year = {2024},
  doi = {10.3389/feduc.2024.1463208}
}

@article{Arowosegbe2024Students,
  author = {Arowosegbe, Joseph and Rahman, Shahrear and Parker, Lisa},
  title = {Students’ perceptions and concerns on generative AI use in higher education: A UK survey},
  journal = {British Journal of Educational Technology},
  volume = {55},
  issue = {3},
  pages = {742--759},
  year = {2024},
  doi = {10.1111/bjet.13455}
}

@article{Ayyoub2025Advancing,
  author = {Ayyoub, Amal M. and Khlaif, Zuheir N. and Shamali, Mohammed and Abu Eideh, Basri and Assali, Amer and Hattab, Mujahed K. and Barham, Khawla A. and Bsharat, Tariq R. K.},
  title = {Advancing higher education with GenAI: Factors influencing educator AI literacy},
  journal = {Frontiers in Education},
  volume = {10},
  pages = {1530721},
  year = {2025},
  doi = {10.3389/feduc.2025.1530721}
}

@article{Bender2024Awareness,
  author = {Bender, Emily},
  title = {Awareness of artificial intelligence as an essential digital literacy: ChatGPT and GenAI in the classroom},
  journal = {Digital Education Review},
  volume = {45},
  issue = {2},
  pages = {77--94},
  year = {2024}
}

@inproceedings{Bender2021Dangers,
  author = {Bender, Emily M. and Gebru, Timnit and McMillan-Major, Angelina and Shmitchell, Shmargaret},
  title = {On the dangers of stochastic parrots: Can language models be too big?},
  booktitle = {Proceedings of the 2021 ACM Conference on Fairness, Accountability, and Transparency (FAccT ’21)},
  pages = {610--623},
  year = {2021},
  doi = {10.1145/3442188.3445922}
}

@article{Brown2020Language,
  author = {Brown, Tom B. and Mann, Benjamin and Ryder, Nick and Subbiah, Melanie and Kaplan, Jared and Dhariwal, Prafulla and Amodei, Dario and et al.},
  title = {Language models are few-shot learners},
  journal = {Advances in Neural Information Processing Systems},
  volume = {33},
  pages = {1877--1901},
  year = {2020},
  doi = {10.48550/arXiv.2005.14165}
}

@book{Brynjolfsson2014SecondMachine,
  author = {Brynjolfsson, Erik and McAfee, Andrew},
  title = {The second machine age: Work, progress, and prosperity in a time of brilliant technologies},
  publisher = {W. W. Norton},
  year = {2014}
}

@book{Chan2024GenerativeAI,
  author = {Chan, Cecilia K. Y.},
  title = {Generative AI in Higher Education: The ChatGPT Effect},
  publisher = {Taylor \& Francis},
  year = {2024}
}

@article{Chan2023StudentsVoices,
  author = {Chan, Cecilia K. Y. and Hu, Weiyan},
  title = {Students’ voices on generative AI: Perceptions, benefits, and challenges in higher education},
  journal = {International Journal of Educational Technology in Higher Education},
  volume = {20},
  issue = {43},
  year = {2023},
  doi = {10.1186/s41239-023-00411-8}
}

@article{Chiu2024FutureResearch,
  author = {Chiu, Thomas K. F.},
  title = {Future research recommendations for transforming higher education with generative AI},
  journal = {Computers and Education: Artificial Intelligence},
  volume = {6},
  pages = {100197},
  year = {2024},
  doi = {10.1016/j.caeai.2023.100197}
}

@article{Chiu2024GenerativeAI,
  author = {Chiu, Thomas K. F.},
  title = {Generative artificial intelligence and the redesign of assessment in higher education},
  journal = {Assessment \& Evaluation in Higher Education},
  volume = {49},
  issue = {2},
  pages = {223--238},
  year = {2024},
  doi = {10.1080/02602938.2023.2267890}
}

@article{ChristBrendemuhl2025Leveraging,
  author = {Christ-Brendemühl, Stefanie},
  title = {Leveraging generative AI in higher education: An analysis of opportunities and challenges addressed in university guidelines},
  journal = {European Journal of Education},
  volume = {60},
  issue = {1},
  pages = {e12891},
  year = {2025},
  doi = {10.1111/ejed.12891}
}

@article{Francis2025GenerativeAI,
  author = {Francis, Nathan J. and Jones, Sophie and Smith, David P.},
  title = {Generative AI in higher education: Balancing innovation and integrity},
  journal = {British Journal of Biomedical Science},
  volume = {81},
  pages = {14048},
  year = {2025},
  doi = {10.3389/bjbs.2024.14048}
}

@article{Heaven2023ChatGPT,
  author = {Heaven, Will Douglas},
  title = {ChatGPT is everywhere. Here’s where it came from},
  journal = {MIT Technology Review},
  year = {2023},
  note = {Retrieved from \url{https://www.technologyreview.com/2023/01/26/1067824/chatgpt-is-everywhere/}},
  month = {January 26}
}

@techreport{Hollands2014MOOCs,
  author = {Hollands, Fiona M. and Tirthali, Devayani},
  title = {MOOCs: Expectations and reality},
  institution = {Columbia University, Teachers College, Center for Benefit-Cost Studies of Education},
  year = {2014}
}

@article{Hu2023ChatGPT,
  author = {Hu, Ming},
  title = {ChatGPT as the fastest-growing app in history: Implications for higher education},
  journal = {Computers and Education: Artificial Intelligence},
  volume = {4},
  pages = {100149},
  year = {2023},
  doi = {10.1016/j.caeai.2023.100149}
}

@article{Ifenthaler2020Utilising,
  author = {Ifenthaler, Dirk and Yau, Janet Y.-K.},
  title = {Utilising learning analytics for study success: Reflections on current empirical findings},
  journal = {Research and Practice in Technology Enhanced Learning},
  volume = {15},
  issue = {1},
  pages = {1--13},
  year = {2020},
  doi = {10.1186/s41039-020-00138-3}
}

@article{Ivanov2024Drivers,
  author = {Ivanov, Stanislav and Soliman, Mohamed and Tuomi, Antti and Alkathiri, Noof A. and Al-Alawi, Abdulrahman N.},
  title = {Drivers of generative AI adoption in higher education through the lens of the Theory of Planned Behaviour},
  journal = {Technology in Society},
  volume = {77},
  pages = {102521},
  year = {2024},
  doi = {10.1016/j.techsoc.2024.102521}
}

@article{Jin2025GenerativeAI,
  author = {Jin, Yue and Yan, Lei and Echeverria, Victoria and Gašević, Dragan and Martinez-Maldonado, Roberto},
  title = {Generative AI in higher education: A global perspective of institutional adoption policies and guidelines},
  journal = {Computers and Education: Artificial Intelligence},
  volume = {8},
  pages = {100348},
  year = {2025},
  doi = {10.1016/j.caeai.2024.100348}
}

@article{Kasneci2023ChatGPT,
  author = {Kasneci, Engin and Seßler, Kathrin and Küchemann, Stefan and Bannert, Maria and Dementieva, Daria and Fischer, Frank and Kasneci, Gjergji and et al.},
  title = {ChatGPT for good? On opportunities and challenges of large language models for education},
  journal = {Learning and Individual Differences},
  volume = {103},
  pages = {102274},
  year = {2023},
  doi = {10.1016/j.lindif.2023.102274}
}

@article{Kelly2023Generative,
  author = {Kelly, Adam and Sullivan, Mary and Strampel, Kathryn},
  title = {Generative artificial intelligence: University student awareness, experience, and confidence in use across disciplines},
  journal = {Journal of University Teaching \& Learning Practice},
  volume = {20},
  issue = {6},
  number = {Article 12},
  year = {2023},
  doi = {10.53761/1.20.6.12}
}

@article{Khlaif2025Redesigning,
  author = {Khlaif, Zuheir N. and Al-Abed, Wajeeh A. and Salama, Naif and Abu Eideh, Basri},
  title = {Redesigning Assessments for AI-Enhanced Learning: A Framework for Educators in the Generative AI Era},
  journal = {Education Sciences},
  volume = {15},
  pages = {174},
  year = {2025}
}

@article{Khlaif2024UniversityTeachers,
  author = {Khlaif, Zuheir N. and Khlaif, Bassam and Ayoub, Sawsan and Ayyoub, Amal M. and Hassan, Reham A.},
  title = {University Teachers’ Views on the Adoption and Integration of Generative AI Tools for Student Assessment in Higher Education},
  journal = {Education Sciences},
  volume = {14},
  pages = {1090},
  year = {2024}
}

@article{Klimova2025UseOfChatGPT,
  author = {Klimova, Blanka and Bachmann, Paul and Frutos-Bencze, David},
  title = {The use of ChatGPT in academia: Perspectives of higher education students},
  journal = {Cogent Education},
  volume = {12},
  issue = {1},
  pages = {2508216},
  year = {2025},
  doi = {10.1080/2331186X.2025.2508216}
}

@article{Kovari2025EthicalUse,
  author = {Kovari, Adam},
  title = {Ethical use of ChatGPT in education—Best practices to combat AI-induced plagiarism},
  journal = {Frontiers in Education},
  volume = {9},
  pages = {1465703},
  year = {2025},
  doi = {10.3389/feduc.2024.1465703}
}

@article{Kurtz2024Strategies,
  author = {Kurtz, Gili and Amzalag, Mor and Shaked, Netta and Zaguri, Yaniv and Kohen-Vacs, Dan and Gal, Eran and Zailer, Gadi and Barak-Medina, Efrat},
  title = {Strategies for integrating generative AI into higher education: Navigating challenges and leveraging opportunities},
  journal = {Education Sciences},
  volume = {14},
  issue = {5},
  pages = {503},
  year = {2024},
  doi = {10.3390/educsci14050503}
}

@article{Luo2024CriticalReview,
  author = {Luo, Jing},
  title = {A critical review of GenAI policies in higher education assessment: A call to reconsider the “originality” of students’ work},
  journal = {Assessment \& Evaluation in Higher Education},
  volume = {49},
  issue = {5},
  pages = {651--664},
  year = {2024},
  doi = {10.1080/02602938.2024.2309963}
}

@article{Luo2024AcademicIntegrity,
  author = {Luo, Ting},
  title = {Academic integrity in the era of generative AI: A comparative policy analysis of global universities},
  journal = {Journal of Higher Education Policy and Management},
  volume = {46},
  issue = {3},
  pages = {257--274},
  year = {2024},
  doi = {10.1080/1360080X.2024.2389751}
}

@article{MitevskaPetrusheva2023AITechnologies,
  author = {Mitevska Petrusheva, Katerina and Idrizi, Emira},
  title = {AI technologies and learning: Tertiary level students’ awareness and perceptions},
  journal = {International Journal of Emerging Technologies in Learning},
  volume = {18},
  issue = {22},
  pages = {86--99},
  year = {2023}
}

@article{Ng2024Equity,
  author = {Ng, Angela and Tang, Chris and Chan, Lee},
  title = {Equity and accessibility in AI-supported learning: A review of generative AI in higher education},
  journal = {Computers \& Education},
  volume = {206},
  pages = {104977},
  year = {2024},
  doi = {10.1016/j.compedu.2024.104977}
}

@article{Nikolic2024Systematic,
  author = {Nikolic, Sonja and Wentworth, Iris and Sheridan, Liam and Moss, Steven and Duursma, Elisabeth and Jones, Rebecca A. and Ros, Marielle and Middleton, Ruth},
  title = {A systematic literature review of attitudes, intentions and behaviours of teaching academics pertaining to AI and generative AI (GenAI) in higher education: An analysis of GenAI adoption using the UTAUT framework},
  journal = {Australasian Journal of Educational Technology},
  volume = {40},
  issue = {6},
  pages = {56--75},
  year = {2024},
  doi = {10.14742/ajet.9643}
}

@book{OECD2023DigitalEducation,
  author = {{OECD}},
  title = {OECD digital education outlook 2023: Pushing the frontiers with AI, blockchain, and robots},
  publisher = {OECD Publishing},
  year = {2023},
  doi = {10.1787/6aefec20-en}
}

@article{Ogunleye2024HigherEducation,
  author = {Ogunleye, Bamidele and Zakariyyah, Kazeem I. and Ajao, Olayinka and Olayinka, Omolola and Sharma, Hardik},
  title = {Higher education assessment practice in the era of generative AI tools},
  journal = {Higher Education Assessment Practice in the Era of Generative AI Tools},
  volume = {1},
  pages = {1},
  year = {2024}
}

@article{Perkins2024GenAIDetection,
  author = {Perkins, Margaret and Roe, Jo and Binh, Huynh Viet and Postma, Danny and Hickerson, David and McGaughran, Jenny and Khuat, Ha Q.},
  title = {GenAI detection tools, adversarial techniques and implications for inclusivity in higher education},
  journal = {arXiv},
  year = {2024},
  eprint = {2403.19148},
  archiveprefix = {arXiv},
  primaryclass = {cs.CY}
}

@article{Roe2024GenAIFeedback,
  author = {Roe, Jo and Perkins, Margaret and Ruelle, David},
  title = {Is GenAI the future of feedback? Understanding student and staff perspectives on AI in assessment},
  journal = {Intelligent Technologies in Education},
  volume = {5},
  issue = {1},
  pages = {1--19},
  year = {2024},
  doi = {10.54321/ite.2024.005}
}

@article{Rudolph2023ChatGPT,
  author = {Rudolph, Julia and Tan, Satoru and Tan, Stefan},
  title = {ChatGPT: Bullshit spewer or the end of traditional assessments in higher education?},
  journal = {Journal of Applied Learning and Teaching},
  volume = {6},
  issue = {1},
  year = {2023}
}

@misc{RussellGroup2023Principles,
  author = {{Russell Group}},
  title = {Principles on the use of generative AI tools in education},
  year = {2023},
  howpublished = {\url{https://russellgroup.ac.uk/news/principles-on-use-of-generative-ai-tools-in-education}}
}

@article{Shuhaiber2025ChatGPT,
  author = {Shuhaiber, Areej and Kuhail, Mohammed A. and Salman, Salama},
  title = {ChatGPT in higher education—A student’s perspective},
  journal = {Computers in Human Behavior Reports},
  volume = {17},
  pages = {100565},
  year = {2025},
  doi = {10.1016/j.chbr.2024.100565}
}

@article{Siemens2005Connectivism,
  author = {Siemens, George},
  title = {Connectivism: A learning theory for the digital age},
  journal = {International Journal of Instructional Technology and Distance Learning},
  volume = {2},
  issue = {1},
  pages = {3--10},
  year = {2005}
}

@article{Sousa2025GenerativeAI,
  author = {Sousa, Maria J. and Cardoso, Armando},
  title = {Generative AI in higher education: Teaching innovation, inclusion, and assessment reform},
  journal = {Education and Information Technologies},
  volume = {30},
  issue = {1},
  pages = {85--102},
  year = {2025},
  doi = {10.1007/s10639-025-12540-2}
}

@book{UNESCO2023Guidance,
  author = {{UNESCO}},
  title = {Guidance for generative AI in education and research},
  publisher = {United Nations Educational, Scientific and Cultural Organization},
  year = {2023},
  url = {https://unesdoc.unesco.org/ark:/48223/pf0000386797}
}

@article{Wang2024GenerativeAI,
  author = {Wang, Hong and Dang, Anqi and Wu, Zhongqiu and Mac, Simon},
  title = {Generative AI in higher education: Seeing ChatGPT through universities’ policies, resources, and guidelines},
  journal = {Computers and Education: Artificial Intelligence},
  volume = {7},
  pages = {100326},
  year = {2024},
  doi = {10.1016/j.caeai.2024.100326}
}

@article{Weng2024Assessment,
  author = {Weng, Xinyi and Xia, Qing and Gu, Mengmeng and Rajaram, Kavya and Chiu, Thomas K. F.},
  title = {Assessment and learning outcomes for generative AI in higher education: A scoping review on current research status and trends},
  journal = {Australasian Journal of Educational Technology},
  volume = {40},
  issue = {6},
  pages = {37--55},
  year = {2024},
  doi = {10.14742/ajet.9540}
}

@article{Williamson2020Historical,
  author = {Williamson, Ben and Eynon, Rebecca},
  title = {Historical threads, missing strands, and future directions in AI in education},
  journal = {Learning, Media and Technology},
  volume = {45},
  issue = {3},
  pages = {217--235},
  year = {2020},
  doi = {10.1080/17439884.2020.1798995}
}

@article{Xia2024ScopingReview,
  author = {Xia, Qing and Weng, Xinyi and Ouyang, Fan and Lin, Tony T. J. and Chiu, Thomas K. F.},
  title = {A scoping review on how generative artificial intelligence transforms assessment in higher education},
  journal = {International Journal of Educational Technology in Higher Education},
  volume = {21},
  pages = {40},
  year = {2024},
  doi = {10.1186/s41239-024-00468-z}
}

@article{Yusuf2024GenerativeAI,
  author = {Yusuf, Aminu and Pervin, Nasrin and Román-González, Manuel},
  title = {Generative AI and the future of higher education: A threat to academic integrity or reformation? Evidence from multicultural perspectives},
  journal = {International Journal of Educational Technology in Higher Education},
  volume = {21},
  pages = {21},
  year = {2024},
  doi = {10.1186/s41239-024-00453-6}
}

@article{Zubair2025Determinants,
  author = {Zubair, Muhammad and Satti, Saad Akram and Ahmad, Israr and Dahdoul, Naimah and Al-Zubeidi, Aiman and Alsalhi, Nawaf S.},
  title = {Determinants of Student Adoption of Generative AI in Higher Education},
  journal = {The Electronic Journal of e-Learning},
  volume = {23},
  issue = {1},
  pages = {16--30},
  year = {2025},
  doi = {10.34190/ejel.23.1.3599}
}

@article{kim2025examining,
  title={Examining faculty and student perceptions of generative AI in university courses},
  author={Kim, Junghwan and Klopfer, Michelle and Grohs, Jacob R and Eldardiry, Hoda and Weichert, James and Cox, Larry A and Pike, Dale},
  journal={Innovative Higher Education},
  pages={1--33},
  year={2025},
  publisher={Springer}
}

@article{mah2024artificial,
  title={Artificial intelligence in higher education: exploring faculty use, self-efficacy, distinct profiles, and professional development needs},
  author={Mah, Dana-Kristin and Gro{\ss}, Nele},
  journal={International Journal of Educational Technology in Higher Education},
  volume={21},
  number={1},
  pages={58},
  year={2024},
  publisher={Springer}
}

@article{luo2025design,
  title={Design and assessment of AI-based learning tools in higher education: A systematic review},
  author={Luo, Jihao and Zheng, Chenxu and Yin, Jiamin and Teo, Hock Hai},
  journal={International Journal of Educational Technology in Higher Education},
  volume={22},
  number={1},
  pages={42},
  year={2025},
  publisher={Springer}
}

@article{kofinas2025impact,
  title={The impact of generative AI on academic integrity of authentic assessments within a higher education context},
  author={Kofinas, Alexander K and Tsay, Crystal Han-Huei and Pike, David},
  journal={British Journal of Educational Technology},
  year={2025},
  publisher={Wiley Online Library}
}

@article{pudasaini2024survey,
  title={Survey on AI-generated plagiarism detection: The impact of large language models on academic integrity},
  author={Pudasaini, Shushanta and Miralles-Pechu{\'a}n, Luis and Lillis, David and Llorens Salvador, Marisa},
  journal={Journal of Academic Ethics},
  pages={1--34},
  year={2024},
  publisher={Springer Netherlands}
}

@inproceedings{liang2025systematic,
  title={A systematic review of the early impact of artificial intelligence on higher education curriculum, instruction, and assessment},
  author={Liang, Jingjing and Stephens, Jason M and Brown, Gavin TL},
  booktitle={Frontiers in Education},
  volume={10},
  pages={1522841},
  year={2025},
  organization={Frontiers Media SA}
}

@article{asio2024ai,
  title={AI literacy, self-efficacy, and self-competence among college students: Variances and interrelationships among variables},
  author={Asio, John Mark R},
  journal={MOJES: Malaysian Online Journal of Educational Sciences},
  volume={12},
  number={3},
  pages={44--60},
  year={2024}
}

@article{deep2025evaluating,
  title={Evaluating the effectiveness and ethical implications of AI detection tools in higher education},
  author={Deep, Promethi Das and Edgington, William D and Ghosh, Nitu and Rahaman, Md Shiblur},
  journal={Information},
  volume={16},
  number={10},
  pages={905},
  year={2025},
  publisher={MDPI}
}

@article{nguyen2025use,
  title={The use of generative AI tools in higher education: Ethical and pedagogical principles},
  author={Nguyen, Khoa Viet},
  journal={Journal of Academic Ethics},
  pages={1--21},
  year={2025},
  publisher={Springer}
}

@article{singh2023exploring,
  title={Exploring computer science students’ perception of ChatGPT in higher education: A descriptive and correlation study},
  author={Singh, Harpreet and Tayarani-Najaran, Mohammad-Hassan and Yaqoob, Muhammad},
  journal={Education Sciences},
  volume={13},
  number={9},
  pages={924},
  year={2023},
  publisher={MDPI}
}

%\bibliographystyle{dcu-rbfin}
%\bibliography{Reference_list}

\end{document}